\documentclass[fleqn,usenatbib]{mnras}
\usepackage{mathtools}
\usepackage{enumitem}
\usepackage{appendix}
\usepackage{xcolor}
\usepackage{subfigure}
\usepackage{graphicx}
\usepackage{amsmath}
\usepackage{figsize}
\usepackage{txfonts}
\usepackage[T1]{fontenc}
\usepackage{threeparttable}
\usepackage{CJKutf8}

\newcommand{\beam}{\mathrm{\mu Jy/beam\cdot km/s}}
\newcommand{\cbeam}{\mathrm{\mu Jy/beam}}
\newcommand{\mbeam}{\mathrm{mJy/beam\cdot km/s}}
\newcommand{\kms}{\mathrm{km\ s^{-1}}}
\newcommand{\jbk}{\mathrm{Jy/beam\cdot km\ s^{-1}}}
\newcommand{\jk}{\mathrm{Jy\cdot km\ s^{-1}}}
\newcommand{\msun}{\mathrm{M_\odot}}
\newcommand{\lsun}{\mathrm{L_\odot}}
\newcommand{\ergs}{\mathrm{\ erg\ s^{-1}}}
\newcommand{\per}[2]{\mathrm{{#1}^{#2}}}
\newcommand{\red}[1]{\textcolor{black}{#1}}

\newcommand{\tnr}[2]{{#1}_\mathrm{#2}}
\newcommand{\sfr}{\msun\ \rm yr^{-1}}

\DeclareRobustCommand{\VAN}[3]{#2}
\let\VANthebibliography\thebibliography
\def\thebibliography{\DeclareRobustCommand{\VAN}[3]{##3}\VANthebibliography}

\title[Revisiting the Dragonfly Galaxy II]{Revisiting the Dragonfly Galaxy II. Young, radiatively efficient radio-loud AGN drives massive molecular outflow in a starburst merger at $z=1.92$}

\author[Yuxing Zhong et al.]{Yuxing Zhong$^{1}$ \begin{CJK*}{UTF8}{gkai}(仲宇星)\end{CJK*},
Akio K. Inoue$^{1,2}$,
Yuma Sugahara$^{2,3}$,
Kana Morokuma-Matsui$^{4}$,
Shinya Komugi$^{5}$,
\newauthor{Hiroyuki Kaneko$^{6,7,3}$,
and Yoshinobu Fudamoto$^{8,2,3}$}
\\
$^{1}$Department of Pure and Applied Physics, Waseda University, 3-4-1 Okubo, Shinjuku, Tokyo 169-8555, Japan\\
$^{2}$Waseda Research Institute of Science and Engineering, Waseda University, 3-4-1, Okubo, Shinjuku, Tokyo 169-8555, Japan\\
{$^{3}$National Astronomical Observatory of Japan, 2-21-1 Osawa, Mitaka, Tokyo 181--8588, Japan}\\
{$^{4}$Center for Computational Sciences, University of Tsukuba, Ten-nodai, 1-1-1 Tsukuba, Ibaraki 305-8577, Japan}\\
{$^{5}$Department of Liberal Arts, Kogakuin University, 2665-1 Nakano-cho, Hachioji, Tokyo 192-0015, Japan}\\
{$^{6}$Joetsu University of Education, 1, Yamayashiki-machi, Joetsu, Niigata 943--8512, Japan}\\
{$^{7}$Ibaraki University, 2-1-1 Bunkyo, Mito, Ibaraki 310-8512, Japan}\\
{$^{8}$Chiba University, 1-33, Yayoicho, Inage-ku, Chiba-shi, Chiba, 263-8522 Japan}}

\date{Accepted XXX. Received YYY; in original form ZZZ}

\pubyear{2023}

\begin{document}
\label{firstpage}
\pagerange{\pageref{firstpage}--\pageref{lastpage}}
\maketitle

\begin{abstract}
Radio-loud active galactic nuclei (RLAGNs) are a unique AGN population and were thought to be preferentially associated with supermassive black holes (SMBHs) at low accretion rates. 
They could impact the host galaxy evolution by expelling cold gas through the jet-mode feedback. 
In this work, we studied CO(6-5) line emission and continuum emission in a high-redshift radio galaxy, MRC 0152-209, at $z=1.92$ using ALMA up to a $0.024\arcsec$-resolution (corresponding to $\sim200$ pc \red{at $z=1.92$}).
This system is a starburst major merger comprising two galaxies: the northwest (NW) galaxy hosting the RLAGN with jet kinetic power $\tnr{L}{jet}\gtrsim2\times10^{46}\ \ergs$ and \red{the other galaxy to the southeast (SE)}.
Based on the SED fitting for the entire system (NW+SE galaxies), we find an AGN bolometric luminosity $\tnr{L}{AGN,bol}\sim3\times10^{46}\ \ergs$ with a lower limit of $\sim0.9\times10^{46}\ \ergs$ for the RLAGN.
We estimate the black hole mass through $M_{\rm BH}-M_\star$ scaling relations and find an Eddington ratio of $\tnr{\lambda}{Edd}\sim0.07-4$ conservatively \red{by adopting the lower limit of $\tnr{L}{AGN,bol}$ and considering the dispersion of the scaling relation}.
These results suggest that the RLAGN is radiatively efficient and the powerful jets could be launched from a super-Eddington accretion disc.
ALMA Cycle 6 observations further reveal a massive ($\tnr{M}{H_2}=(1.1-2.3)\times10^9\ \rm M_\odot$), compact ($\sim500$ pc), and \red{monopolar} molecular outflow perpendicular to the jet axis.
The corresponding mass outflow rate \red{($1200^{+300}_{-300}-2600^{+600}_{-600}\ \sfr$)} is comparable with the star formation rate of at least $\sim2100\ \sfr$.
\red{Depending on the outflowing molecular gas mass}, the outflow kinetic power/$\tnr{L}{AGN,bol}$ ratio of $\sim0.008-0.02$, and momentum boost factor of $\sim3-24$ agree with a radiative-mode AGN feedback scenario.
On the other hand, the jets can also drive the molecular outflow within its lifetime of $\sim2\times10^5$ yr without additional energy supply from AGN radiation.
The jet-mode feedback is then capable of removing all cold gas from the host galaxy through the long-term, episodic launching of jets.
Our study reveals a unique object where starburst activity, powerful jets, and rapid BH growth co-exist, which may represent a fundamental stage of AGN-host galaxy co-evolution.
\end{abstract}


\section{Introduction}\label{intro}
RLAGNs are rare amongst all AGN populations ($15-20\%$; \citealt{1989AJ.....98.1195K,2018MNRAS.475.3429W}).
At higher redshifts ($1<z<2$), there is an increasing space density of RLAGNs hosted by radiatively efficient (Eddington ratio $\tnr{\lambda}{Edd}\gtrsim3\times10^{-2}$; \citealt{2008MNRAS.388.1011M}) SMBHs linked to high-excitation radio galaxies (see \citealt{2020NewAR..8801539H} and references therein).
These high-\textit{z} galaxies hosting RLAGNs are often referred to as high-redshift radio galaxies (HzRGs) with rest-frame radio power $\tnr{L}{500\ MHz}>10^{27.5}$ W $\per{Hz}{-1}$ \citep{2008A&ARv..15...67M}.
The typical infrared (IR) luminosity of HzRGs is found to exceed $\rm 10^{12}\ \lsun$, in agreement with the classifications of ultra-luminous infrared galaxies (ULIRG; $\tnr{L}{IR}\geq10^{12}\ \lsun$), while some are found to be even brighter than $10^{13}\ \lsun$, entering hyper-LIRG (HyLIRG; $\tnr{L}{IR}\geq10^{13}\ \lsun$) regime \citep{2014AA...566A..53D}.
U/HyLIRG populations are often massive, starburst systems and are thought to be triggered by galaxy mergers \citep{1996ARA&A..34..749S}.
\red{The fact that HzRGs are also U/HyLIRGs suggests that powerful RLAGNs are often associated with major mergers, which is evident by accumulating observations \citep[e.g.,][]{2012MNRAS.419..687R,2015ApJ...806..147C,2022MNRAS.510.1163P,2023MNRAS.522.1736P}}.
\red{An investigation into the cold molecular gas -- the raw fuel that feeds both the growth of SMBHs and star formation -- provides insights to enhance our understanding of galaxy mergers, starburst galaxies (SBGs), and AGN/SMBHs as the basis for co-evolution in the high-\textit{z} universe.}

Recent studies of HzRGs find that the most powerful RLAGNs are likely to be associated with AGN with bolometric luminosity ($\tnr{L}{AGN,bol}$) comparable to high-\textit{z} QSOs as the SMBHs are in a fast growth phase \citep[e.g.,][]{2017A&A...600A.121N,2021ApJ...921...51I}.
For QSOs, the accreting gas around the SMBHs usually forms an optically thick, geometrically thin accretion disc \citep{1973A&A....24..337S,1998tbha.conf..148N}.
The question then arises, that, a standard thin disc ($0.01\lesssim\tnr{\lambda}{Edd}\lesssim1$) is widely considered jet phobic, possibly because the large-scale magnetic field that powers the radio jets is diffused out of the accretion disc faster than being dragged inward \citep{1994MNRAS.267..235L,2012MNRAS.424.2097G} \citep[\red{however, recent simulations show that thin discs can sustain large-scale poloidal magnetic fields around the BH; see, e.g.,}][]{2008ApJ...677.1221R,2019MNRAS.487..550L}.
One possible explanation for the origin of the powerful jets is that these most powerful HzRGs (jet kinetic power $\tnr{L}{jet}\gtrsim10^{47}$ erg $\per{s}{-1}$ and $\tnr{L}{AGN,bol}\gtrsim10^{46}$ erg $\per{s}{-1}$) may in general host over-massive SMBHs ($10^{-2}\lesssim\tnr{M}{BH}/\tnr{M}{\star}\lesssim2\times10^{-1}$; \citealt{2021ApJ...921...51I}) that lie far above the BH mass $(M_\text{BH})$ - stellar mass ($M_\star$) scaling relations.
The $\tnr{M}{BH}$ estimated based on the scaling relations is underestimated, thus the Eddington ratio corresponding to a standard thin disc is an overestimation.
These SMBHs are actually surrounded by optically thin, geometrically thick sub-Eddington accretion discs \citep{2021ApJ...921...51I}.
However, using rest-frame optical emission lines (H$\alpha$, H$\beta$, Mg {\small II}, and C {\small IV}), \citet{2023A&A...672A.164P} estimated the BH masses of a sample of RLAGNs at $0.3<z<4$ with $\tnr{L}{jet}<10^{47}$ erg $\per{s}{-1}$. 
They \red{found a good agreement of the $M_\text{BH}$ - $M_\star$ scaling relations of RLAGNs and other AGN populations in samples across different redshifts.}
Therefore, these HzRGs seem to follow the scaling relations statistically.
This raises another possibility, that is, many HzRGs host SMBHs accreting above the Eddington limit, and thus the accretion discs are both optically and geometrically thick, capable of powering the jets \citep{2015ASSL..414...45T}.

AGN feedback has been an important topic to study AGN-host galaxy co-evolution because it may lead to an outflow (compression) of the gas to suppress (enhance) the host galaxy star formation \citep[e.g.,][]{2012A&A...537L...8C,2014A&A...562A..21C,2019ApJ...881..147S,2023MNRAS.522.4548D}.
Outflows in ionized and/or molecular forms are often observed in galaxies hosting AGNs with $\tnr{L}{AGN,bol}\gtrsim10^{44}$ erg $\per{s}{-1}$ from local to high redshifts \citep[e.g.,][]{2017A&A...601A.143F,2019MNRAS.483.4586F,2019A&A...630A..59B}, \red{as well as those with lower AGN luminosities \citep{2021MNRAS.503.1780J}.}
These outflows are often ascribed to the radiative-mode AGN feedback no matter it is through the radiation pressure on dust or through shocks generated from inner AGN winds \citep{2014MNRAS.444.2355C,2018MNRAS.476..512I}.
This radiative-mode AGN feedback has the potential to remove a significant fraction of molecular gas from the AGN host galaxy \citep{2012ApJ...745L..34Z}.
Without molecular gas as the fuel to support ongoing star-forming activities, \red{the SFR can decrease and these galaxies will finally become quiescent.}
In addition to the radiative-mode, the jet (kinetic)-mode AGN feedback, through the couplings between jets and the host galaxy interstellar medium (ISM), has also been found to be efficient in accelerating gas with colossal energy injections to power the outflow \citep{2012ApJ...757..136W,2018MNRAS.479.5544M,2022MNRAS.516..766M}.
Both simulations and observations find that the jet-mode feedback can lead to the quenching of host galaxies \citep{2006ApJ...650..693N,2024MNRAS.527.5988H}.
Because of the simultaneously high $\tnr{L}{jet}$ and $\tnr{L}{AGN,bol}$ in powerful HzRGs, jet- and radiative-mode feedback co-exist in the host galaxies \citep{2024MNRAS.527.5988H}.
This makes the outflow mechanism mysterious: which is the main driver of the outflow?
Additionally, do outflows in these HzRGs show distinctly different molecular outflow properties when compared to QSOs and low-\textit{z} AGNs?
These questions remain unclear and require accumulating observational studies to investigate.

RLAGNs at an early evolution phase, at a time of radio jets remaining powerful and efficiently injecting their energy into the ISM \citep{2022MNRAS.516..766M}, are good targets to investigate the jet-mode feedback.
Moreover, by investigating the morphology of RLAGN host galaxies at $z>1$ using HST WFPC3 images, \citet{2015ApJ...806..147C} found that 92\% of these RL galaxies show recent or ongoing merging events.
This finding is suggestive of possible merger-triggered AGN activities.
Therefore, young RLAGNs at high-$z$ serve as ideal proxies to scrutinize the co-evolution of AGN and its host galaxy through processes like galaxy interaction/merging and AGN feedback since the stochastic gas inflows accompanied with these events may both fuel star formation and trigger AGN activities \citep{2020NewAR..8801539H,2021ApJ...923...36S}.

In this work, we study a HzRG -- MRC 0152-209 ($\tnr{L}{147\ MHz}=3.2(\pm1.0)\times10^{28}$ W $\per{Hz}{-1}$), named Dragonfly galaxy, which has starbursts \citep[$\mathrm{SFR\sim3000\ M_\odot\ yr^{-1}}$;][]{2014AA...566A..53D}.
It is a HyLIRG ($\tnr{L}{IR}\sim2\times10^{13}\ \lsun$) and a major merger comprising three components: the North-West (NW) galaxy, the South-East (SE) galaxy, and a possible companion galaxy. 
Its double radio hotspots were identified by the Very Large Array (VLA) at 4.7 and 8.2 GHz \citep{2000AAS..145..121P}.
\citet{2023MNRAS.522.6123Z} (referred to as Paper I) further investigated the radio hotspots combining high-resolution and high-frequency VLA and Atacama Large Millimeter/submillimeter Array (ALMA) observations.
They found that the Dragonfly galaxy can be classified as a Compact-Steep-Spectrum source.
The radio hotspots have an age of $\sim2\times10^5$ yr, in line with the typical order of magnitude of young RLAGNs.
We further present high-angular resolution ALMA observations ($0.08\arcsec$ for Cycle 4 and $0.02\arcsec$ for Cycle 6) of CO(6-5) line emission from the Dragonfly galaxy to investigate the molecular gas within sub-kpc regions.
Our new study provides a unique view of the molecular gas distribution among this merging system, revealing details of the AGN-driven outflow.

Throughout this paper, we assume a $\mathrm{\Lambda CDM}$ cosmology with $\Omega_m=0.309$, $\Omega_\Lambda=0.691$, and $H_0=\mathrm{67.7\ km\ s^{-1}\ Mpc^{-1}}$ \citep{2016A&A...594A..13P}.
Based on these assumptions, the luminosity distance of the Dragonfly galaxy is $\sim$15200 Mpc, and $1\arcsec$ corresponds to a projected physical scale of 8.62 kpc.\footnote{The calculation has made use of \citet{2006PASP..118.1711W}.}

\begin{table*}
\caption{Summary of the Observations}
\label{tab:obs}
\begin{tabular}{cccccccc}
\hline
Observation & Date & Frequency & Band & Beam size & Position Angle & $\sigma$ \\
\hline
ALMA Cycle 4 continuum & 9 and 17th August, 2017 & 237 GHz & Band 6 & $0.11\arcsec\times0.08\arcsec$ & $80^\circ$ & 11 $\cbeam$ \\
ALMA Cycle 4 CO(6-5) & 9 and 17th August, 2017 & 237 GHz & Band 6 & $0.12\arcsec\times0.08\arcsec$ & $78^\circ$ & 183 $\beam$\\
ALMA Cycle 6 continuum & 23th June, 2019 & 237 GHz & Band 6 & $0.026\arcsec\times0.023\arcsec$ & $24^\circ$ & 6.5 $\cbeam$ \\
ALMA Cycle 6 CO(6-5) & 23th June, 2019 & 237 GHz & Band 6 & $0.027\arcsec\times0.024\arcsec$ & $24^\circ$ & 102 $\beam$\\
VLA BnA-configuration & 29th May, 2015 & 44 GHz & Band Q & $0.14\arcsec\times0.08\arcsec$ & $80^\circ$ & 18 $\cbeam$ \\
\hline
\end{tabular}
\end{table*}

\section{Observations}
\subsection{ALMA Band 6}
ALMA Cycle 4 observations \red{(Project ID: 2016.1.01417.S, PI: Bjorn Emonts)} were conducted on 9 and 17 August 2017 for 1.2 hours on-source time with 45 antennas and baselines of 12-m array up to $\sim3.6$ km.
ALMA Cycle 6 observations \red{(Project ID: 2018.1.00293.S, PI: Bjorn Emonts)} were conducted on 23 June 2019 for 2.2 hours on-source time with 48 antennas and baselines of the 12-m array up to 11.5 km.
For both observations, there are four spectral windows configured to cover two 3.75 GHz bands, one of which includes $235.83-239.58$ GHz to observe the redshifted CO(6-5) line emission $(\nu\mathrm{_{rest}=691.47\ GHz})$ and another includes $251.20-254.95$ GHz such that only continuum is observed.
\red{The redshifted frequency is estimated based on $z=1.92$ determined by the observation of CO(1-0) line emission \citep{2011ApJ...734L..25E}.}
The data calibrations were performed via the ALMA pipeline built within CASA (Common Astronomy Software Applications; \citealt{2007ASPC..376..127M,2022PASP..134k4501C}) version 4.7.2 for Cycle 4 and 5.4.0 for Cycle 6, respectively, by running the calibration scripts supplied with the data by the North American ALMA Science Center (NAASC).

Prior to imaging the CO(6-5) line emission, we subtracted the continuum emission in the \textit{uv}-plane.
To do so, we first flagged the channels that include the real line emission, as well as those that include pseudo-line emissions because of the strong atmospheric absorption between 237.15 GHz and 239.1 GHz (private communication with Hiroshi Nagai through ALMA helpdesk, Jul 27, 2022).
We then estimated the continuum emission by a linear function to the line-free channels and subtracted it in the \textit{uv}-plane by using the task \texttt{uvcontsub}.

We adopted the same methodology for Cycle 4 and 6 observations to image the line and continuum emissions in this work.
First, we created a dirty image without any clean to calculate the root-mean-square noise $(\sigma)$ under the `briggs' weighting with a robustness parameter +0.5.
Then, we cleaned the image non-interactively by setting $3\sigma$ as the stop threshold of the cleaning.
We used the `h\"{o}gbom' deconvolution algorithm to produce the restored image and applied a primary beam correction on the restored image.
For the CO(6-5) line emission, the channel width was set to be 20 $\kms$ for \red{both Cycle 4 and 6 observations}, and the reference frequency was set as 236.69 GHz, corresponding to the line at $z=1.9214$.
We further imaged the CO(6-5) line emission using concatenated Cycle 4 and 6 (C46 hereafter) data.
A robustness parameter of $+2.0$ was chosen for the `briggs' weighting and a spectral resolution of 15 $\kms$ was adopted.
We chose this natural weighting to balance the beam size and noise level.
The basic information and properties of the cleaned images of line and continuum emissions are summarized in Table.~\ref{tab:obs}.

The properties of the line profiles and their associated physical properties based on the Cycle 4 and 6 combined data are not presented in this paper due to an abnormal elevation in flux densities.
This was considered a software issue of CASA, which has been reported to the CASA development team (private communication with Hiroshi Nagai through ALMA helpdesk, Mar 25, 2022).
\red{This abnormal elevation only influences the combined dataset and has no impact on Cycle 4 and 6 observations individually.}

\subsection{VLA Band Q}
The VLA observations reported in this work were conducted in BnA-configuration centered at 44 GHz with an effective bandwidth of 7.5 GHz.
The total on-source time is 42 min.

The observations were calibrated by requesting pipeline calibrations through the NRAO Science Helpdesk.
We imaged the radio continuum emission using the `h\"{o}gbom' deconvolution algorithm.
A `briggs' weighting scheme with a robustness parameter +0.5 was chosen to multiply the visibility value during the gridding of the dirty image.
To create a clean image, we put a mask on the strongest signal and manually iterated until the strongest signal reached the noise level in order to circumvent artefacts as a result of over-cleaning
Due to the low signal-to-noise ratio of each spectral window, no self-calibration in any dataset can be applied, leaving low-level sidelobe contamination on the clean images.

\begin{figure*}
\begin{center}
\includegraphics[width=\textwidth]{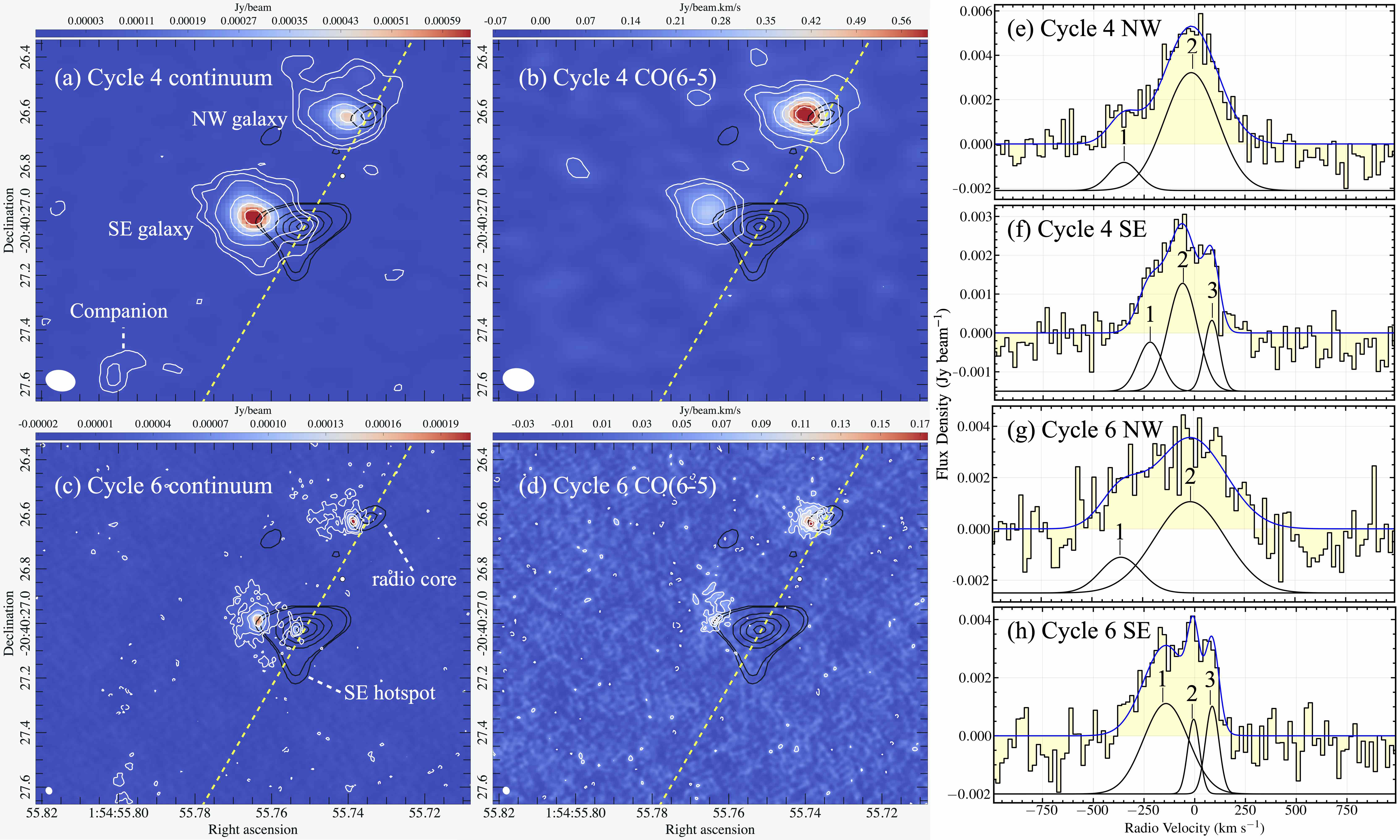}
\caption{A global view of the Dragonfly galaxy.
(a) ALMA Cycle 4 continuum imaging ($\sigma=11\ \cbeam$) with contour levels of $[3, 5, 10, 25, 40, 55]\times\sigma$.
(b) ALMA Cycle 4 CO(6-5) mom0 map ($\sigma=23\ \mbeam$) with contour levels of $[3, 5, 10, 15, 25]\times\sigma$.
(c) ALMA Cycle 6 continuum imaging ($\sigma=6.5\ \cbeam$) with contour levels of $[3, 5, 10, 20, 30]\times\sigma$.
(d) ALMA Cycle 6 CO(6-5) mom0 map ($\sigma=13\ \mbeam$) with contour levels of $[3, 5, 7, 9, 11, 13]\times\sigma$.
Throughout all images, the world coordinates are identical.
The black contours indicate the VLA 44 GHz observation in BnA-configuration ($\sigma=18\ \cbeam$) with contour levels of $[5, 7, 20, 40, 60, 80]\times\sigma$ with the detection coinciding with the NW galaxy indicating the radio core and the detection adjacent to the SE galaxy indicating the SE hotspot.
The yellow dashed line indicates the jet axis that links the double radio hotspot (see also Fig.~1 in Paper I; \citealt{2023MNRAS.522.6123Z}).
The beam size of the corresponding image is marked in the bottom left corner.
In panels (e) - (h), we show line profiles extracted from a circular aperture that encloses contours defined by $3\sigma$.
}
\label{fig:imaging}
\end{center}
\end{figure*}

\section{Results}
\subsection{Morphology}
In panels (a) and (c) of Fig.~\ref{fig:imaging}, we show the images of the dust continuum, and in panels (b) and (d), we show the integrated intensity (mom0) map of CO(6-5) line emission integrated over $-500$ to $+250\ \kms$, for both Cycle 4 and 6 observations.
All line and continuum images and contours are overlaid directly based on the world coordinate system (WCS) after the clean procedure without any manual manipulation to associate the features observed in different bands.
\red{A systematic offset in the WCS of VLA may exist because of the astrometry worsened by the very extended baseline configurations and atmospheric conditions\footnote{https://help.almascience.org/kb/articles/what-is-the-absolute-astrometric-accuracy-of-alma}.
Under a typical condition\footnote{https://science.nrao.edu/facilities/vla/docs/manuals/oss/performance/positional-accuracy}, the VLA and ALMA observations are consistent in positions within $2\sigma$ (see also Paper I, \citealt{2023MNRAS.522.6123Z})}

NW and SE galaxies show individual, putative disc structures in both line and continuum emissions of Cycle 4 observations.
The molecular tidal bridge identified in ALMA Cycle 2 observations between SE and NW galaxies is not observed in both Cycle 4 and 6 observations because of its diffuseness and low molecular gas mass (see \citealt{2015A&A...584A..99E, 2023ApJ...951...73L} for details).
In addition to these two galaxies, the Cycle 4 dust continuum has identified a companion component (see panel (a) in Fig.~\ref{fig:imaging}), which is undetected in both Cycle 4 and 6 mom0 maps.
In Cycle 2 observations, this companion was identified in CO(6-5) line emission while missing in the dust continuum, and argued to be a small companion galaxy \citep{2015A&A...584A..99E}.

The highest resolution Cycle 6 observations have revealed the detailed morphology of the Dragonfly galaxy. 
The CO(6-5) line emission of the NW galaxy is still disc-like but is highly compact with a 2D Gaussian component of $0.091\arcsec(\pm0.01\arcsec)\times0.083\arcsec(\pm0.01\arcsec)$, corresponding to a radius of $\tnr{R}{NW}\sim0.8$ kpc, after being deconvolved from the beam.
This size is more compact than $\tnr{R}{NW}\gtrsim1$ kpc based on Cycle 2 and 4 observations.
\red{To investigate whether such compactness is attributed to the non-detection of a significant fraction of extended molecular gas in Cycle 6 observations}, we compare the integrated intensity $\tnr{I}{CO(6-5)}$ calculated in three cycles (see \S\ref{sec:line profile} and the rightmost panel in Fig.~\ref{fig:imaging}). 
In Cycle 6, the calculated $I_\text{CO(6-5)}=1.79\pm0.13\ \jk$ is consistent with $I_\text{CO(6-5)}=2.0\pm0.2\ \jk$ calculated by \citet{2015A&A...584A..99E} in Cycle 2 and $I_\text{CO(6-5)}=1.98\pm0.13\ \jk$ calculated in Cycle 4 within 1$\sigma$ uncertainty.
This suggests that the bulk of molecular gas is concentrated within a sub-kpc scale disc, though there can be $\sim10$ per cent of the diffuse molecular gas missed by the high-resolution beam.
In the dust continuum, the NW galaxy has its flux density concentrated within a radius of $\sim0.07\arcsec\ (\approx0.6\ \mathrm{kpc})$, agreeing well with the molecular gas distribution, but shows diffuse emissions extended to the east.
As discussed in \S\ref{sec: beads on a string}, this extended feature may originate from the interactions between the two galaxies.

The CO(6-5) line emission distribution of the SE galaxy in Cycle 6 significantly differs from that observed in Cycle 2 and 4 observations, and the zoom-in is shown in Fig.~\ref{fig:se_zoomin}.
The main feature of the SE galaxy is a linear structure (labeled by \#Main) elongated along the southeast to northwest direction.
To the north of the centroid of the SE galaxy, there are several clumps of CO(6-5) line emission (white contours in region \#Tidal), which are in agreement with the distribution of the dust continuum.
\red{These clumps viewed at $\sim100$ pc-scale form a filamentary structure on a $\sim600$ pc-scale and may come into existence through the interactions between two galaxies (see \S\ref{sec: beads on a string} for a detailed discussion).}
On the opposite side of these clumps, there is no line emission detected.
An equivalent lack of emission is seen in the dust continuum as well, save for a sub-component (see next paragraph).
\red{Additionally, in HST NICMOS F160W imaging (Fig.~\ref{fig:hst imaging}), the SE galaxy is associated with a tidal tail extended over 10 kpc.
Such a long tidal tail is commonly observed in wet mergers whose bulge separation is smaller than 5 kpc \citep{2020MNRAS.499.3399R} due to the tidal stripping \citep{2010gfe..book.....M}, that is, the tidal force strips a significant fraction of the molecular gas away from the galaxy.}

The fact that the double radio hotspots identified in VLA 4.7, 8.2, and 44 GHz observations are symmetric relative to the NW galaxy suggests that the RLAGN resides in the NW galaxy (see Fig.~1 in Paper I; \citealt{2023MNRAS.522.6123Z}).
This is supported by the identification of the radio core that overlaps with the NW galaxy (see Fig.~\ref{fig:imaging}), though it is offset from the centroid of ALMA Cycle 6 dust continuum by $0.04\arcsec$. 
The 237 GHz continuum at $0.024\arcsec$-resolution further reveals a sub-component adjacent to and offset from the bulk of SE galaxy dust continuum by $\sim0.15\arcsec$, corresponding to $\sim1.3$ kpc (Fig.~\ref{fig:se_zoomin}).
This sub-component has its location coincided with the SE hotspot with an offset of merely $0.008\arcsec$.
It is then argued to be the radio hotspot originating from the interaction between the medium and radio jet launched by the AGN and its flux density is dominated by the synchrotron radiation \citep[Paper I,][]{2023MNRAS.522.6123Z}.
If the medium at play is the ISM of the SE galaxy, the cool gas can be either blown away by the kinetic energy or heated up to a higher temperature by the thermal energy of the radio jet.
\begin{table*}
\caption{Measured and Derived Physical Properties from the Line Profiles}
\label{tab:lines}
\red{
\begin{tabular}{ccccccccc}
\hline
Component & Observation & Center & FWHM  & $I_\text{CO(6-5)}$ & $L^{\prime}_{\rm CO(6-5)}$ & $\tnr{r}{65/10}$ & $M_\mathrm{H_2}$ &  \\
 & & ($\kms$) & ($\kms$) & $(\jk)$ & $(\mathrm{K\ km\ s^{-1}\ pc^2})$ &  & $(\msun)$ \\
 (1) & (2) & (3) & (4) & (5) & (6) & (7) & (8) \\
\hline
NW - 1 & Cycle 4 CO(6-5) & $-350\pm27$ & $174\pm63$ & $0.23\pm0.08$ & $1.2\pm0.4\times10^9$ & $17-36$ & $1-2.1\times10^{9}$ \\
NW - 2 & Cycle 4 CO(6-5) & $-17\pm9$ & $310\pm22$ & $1.75\pm0.1$ & $9.4\pm0.5\times10^9$ & 13 & $2.1\pm0.1\times10^{10}$ \\
\hline
NW - 1 & Cycle 6 CO(6-5) & $-361\pm22$ & $232\pm52$ & $0.31\pm0.08$ & $1.7\pm0.4\times10^9$ & $17-36$ & $1.3-2.8\times10^{9}$ \\
NW - 2 & Cycle 6 CO(6-5) & $-20$ (fixed) & $414\pm32$ & $1.48\pm0.1$ & $8.0\pm0.5\times10^9$ & 13 & $1.8\pm0.1\times10^{10}$ \\
\hline
SE - 1 & Cycle 4 CO(6-5) & $-220$ & $141\pm37$ & $0.19\pm0.06$ & $1.0\pm0.3\times10^9$ & $17-36$ & $0.8-1.7\times10^{9}$ \\
SE - 2 & Cycle 4 CO(6-5) & $-59\pm13$ & $171\pm40$ & $0.5\pm0.1$ & $2.7\pm0.5\times10^9$ & 13 & $6.0\pm1.2\times10^{9}$ \\
SE - 3 & Cycle 4 CO(6-5) & $86\pm11$ & $89\pm25$ & $0.17\pm0.08$ & $0.9\pm0.4\times10^9$ & 13 & $2.0\pm1.0\times10^{9}$ \\
\hline
SE - 1 & Cycle 6 CO(6-5) & $-141\pm28$ & $260\pm58$ & $0.86\pm0.19$ & $4.6\pm1.0\times10^9$ & 13 & $1.0\pm0.2\times10^{10}$ \\
SE - 2 & Cycle 6 CO(6-5) & $-5\pm13$ & $69\pm33$ & $0.19\pm0.13$ & $1.0\pm0.7\times10^9$ & 13 & $2.3\pm1.5\times10^{9}$ \\
SE - 3 & Cycle 6 CO(6-5) & $86\pm12$ & $78\pm26$ & $0.25\pm0.09$ & $1.3\pm0.5\times10^9$ & 13 & $3.0\pm1.1\times10^{9}$ \\
\hline
SE - \#Main \#1 & Cycle 6 CO(6-5) & $-141\pm32$ & $192\pm66$ & $0.26\pm0.09$ & $1.4\pm0.5\times10^9$ & 13 & $3.1\pm1.1\times10^{9}$ \\
SE - \#Main \#2 & Cycle 6 CO(6-5) & $46\pm18$ & $142\pm32$ & $0.26\pm0.09$ & $1.4\pm0.5\times10^9$ & 13 & $3.1\pm1.1\times10^{9}$ \\
SE - \#Tidal & Cycle 6 CO(6-5) & $-171\pm16$ & $288\pm38$ & $0.39\pm0.04$ & $2.1\pm0.2\times10^9$ & $17-36$ & $1.7-3.6\times10^{9}$ \\
\hline
\end{tabular}}
\begin{tablenotes}
{\item \raggedright 
Column (1): measurements of the components labeled in the line profiles in Fig.~\ref{fig:imaging} and Fig.~\ref{fig:se_zoomin}.
Column (2): component name as labeled in the line profiles of Fig.~\ref{fig:imaging} and Fig.~\ref{fig:se_zoomin}.
Column (3): velocity center, where the zero velocity is set as 236.69 GHz, corresponding to the CO(6-5) line emission at $z=1.9214$.
Column (4): FWHM of CO(6-5) component.
Column (5): the integrated intensity of CO(6-5) line emission.
Column (6): the integrated brightness temperature of CO(6-5) line emission.
Column (7): the line intensity ratio used to convert $\tnr{I}{CO(6-5)}$ to $\tnr{I}{CO(1-0)}$ to estimate the molecular gas mass based on (\citealt{2015A&A...584A..99E}; see texts in \S\ref{sec: gas mass}).
Column (8): the molecular gas mass.
}
\end{tablenotes}
\end{table*}

\subsection{Line Profiles}\label{sec:line profile}
In the third column of Fig.~\ref{fig:imaging}, we show the spectra of NW and SE galaxies observed in Cycle 4 and 6.
\red{For both NW and SE galaxies, the line profiles are extracted using a circular aperture that encloses the spatial area defined by 3$\sigma$-level contours.}
The physical properties measured and derived from the line profiles are listed in Table.~\ref{tab:lines}.

The NW galaxy shows a clear line splitting into two velocity components and a broadening of the main component, and thus the line profile can be fitted by two Gaussian components with velocity center ($\tnr{v}{cen}$), amplitude, and full width at half maximum (FWHM) as fitting parameters.
In Cycle 4, the main component (NW - 2 in Table.~\ref{tab:lines}) has $\tnr{v}{cen}=-17\pm9\ \kms$ and $\mathrm{FWHM=310\pm22\ \kms}$, consistent with the measurements of $\tnr{v}{cen}=-30\pm10\ \kms$ and $\mathrm{FWHM=360\pm20\ \kms}$ in low-resolution Cycle 2 observations \citep{2015A&A...584A..99E}.
In Cycle 6, we first applied Hanning smooth to the spectrum and then adopted a 2-component fitting by fixing $\tnr{v}{cen}$ of the main component (NW - 2) to -20 km s$^{-1}$ found in both Cycle 2 and 4 observations within $1\sigma$ uncertainty.
Otherwise, given the low signal-to-noise ratio, the fitting algorithm returned a third component with uncertainties exceeding 100\%.
The resultant Cycle 6 NW - 2 has an FWHM of $414\pm32\ \kms$, larger than that of Cycle 4 but consistent with Cycle 2 within $1\sigma$ uncertainty.
The blueshifted component (NW - 1) is offset from the main component by at least $300\ \kms$ in all observations.
It has a broader FWHM in Cycle 6 than in Cycle 4 but almost the same amplitude, leading to a slightly higher $\tnr{I}{CO(6-5)}$ by 0.08 $\jk$.
As we will discuss in \S\ref{sec: jet-mode}, this blueshifted component (NW - 1) may represent the outflow driven by the expanding bubble attributed to the radio jet in a 3D shell-like geometry \citep[e.g.,][]{2019A&A...632A..61G}.

\begin{figure*}
\begin{center}
\includegraphics[width=\textwidth]{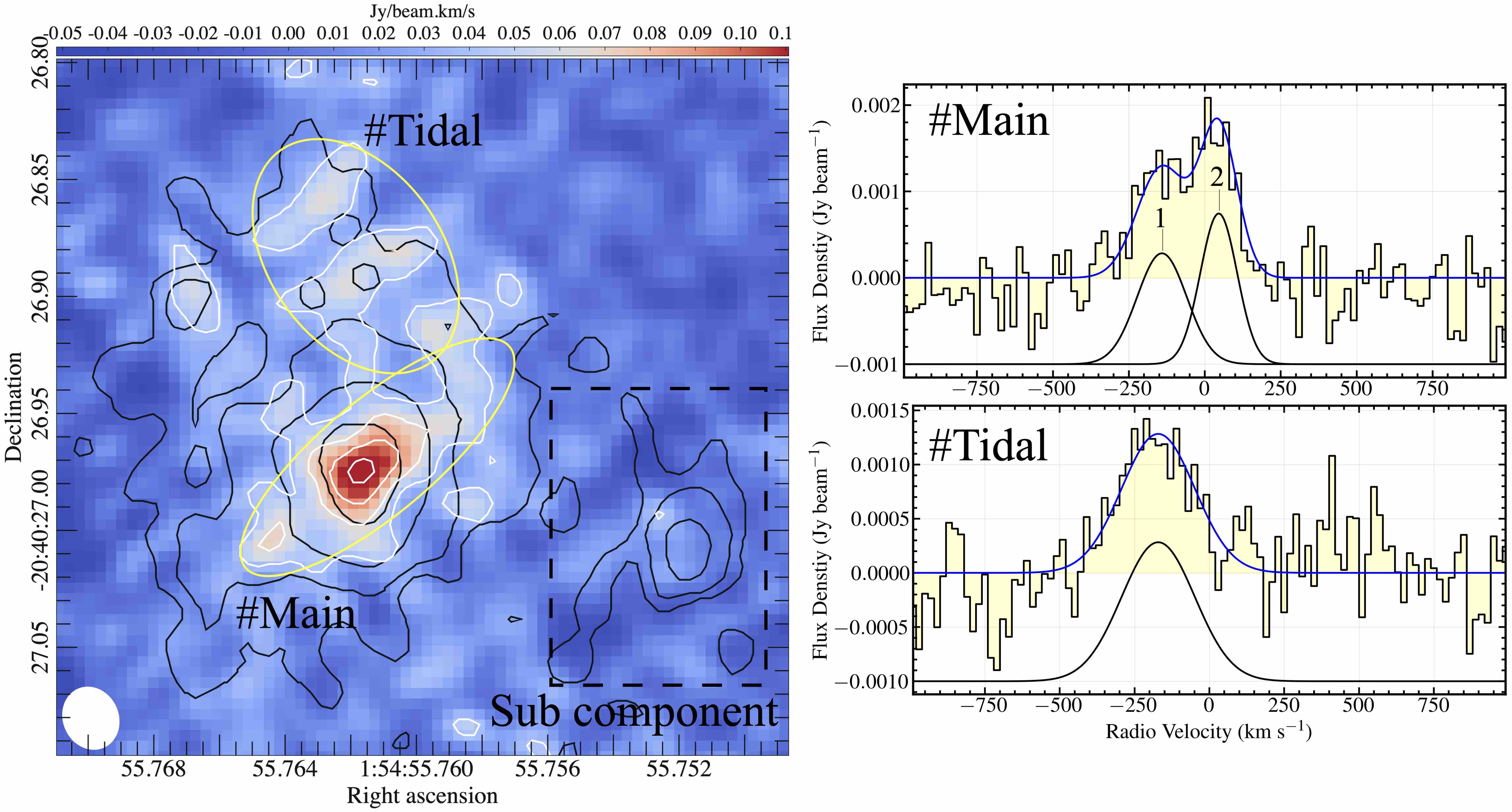}
\caption{\label{fig:se_zoomin}
The zoom-in investigation of the SE galaxy with white contours for CO(6-5) and black contours for 237 GHz continuum laid on the CO(6-5) mom0 map based on Cycle 6 observations.
We divide the mom0 map of the SE galaxy into two regions: \#Main and \#Tidal.
The \#Tidal region is named after its possible association with the tidal bridge (see \S\ref{discus:late-stage merger}).
There is a continuum "sub-component" $\sim1$ kpc to the west of \#Main and coincided with the location of the SE hotspot.
The line profiles extracted from \#Main and \#Tidal regions using the elliptical apertures indicated in the figure are shown in the right column.
The structure in \#Main shows a bimodal line profile characteristic of a rotating structure.
The ellipse in the bottom left corresponds to the beam size.
}
\end{center}
\end{figure*}

The SE galaxy has its line profile decomposable into three Gaussian components in Cycle 4, whereas large uncertainties are left, and the sum of the integrated intensities $\tnr{I}{CO(6-5),SE}=0.86\pm0.14\ \mathrm{Jy/beam\cdot km\ s^{-1}}$ lies far below the value of $1.4\pm0.2\ \mathrm{Jy/beam\cdot km\ s^{-1}}$ observed in Cycle 2 data by \citep{2015A&A...584A..99E}.
In Cycle 6, $\tnr{I}{CO(6-5),SE}$ is $1.3\pm0.37\ \mathrm{Jy/beam\cdot km\ s^{-1}}$, showing no significant deviations from that in the literature, though the SE galaxy shows a much more complex molecular gas distribution in Cycle 6.
Additionally, in Cycle 2 and 4 observations, the brightest component has a center of $\sim60$ km s$^{-1}$ (SE - 2 of Cycle 4 in Table.~\ref{tab:lines}).
However, the blueshifted component (SE - 1 of Cycle 6 in Table.~\ref{tab:lines}) has its $\tnr{I}{CO(6-5)}$ dominant over the other two components in Cycle 6 observations.
To understand the origin of such a change in the brightest component, we zoom in on the SE galaxy and divide its mom0 map into two regions: \#Main and \#Tidal, as shown in the left column of Fig.~\ref{fig:se_zoomin}.
\red{\#Main indicates the main structure in the SE galaxy and the region \#Tidal is named after its possible tidal origin to be discussed in \S\ref{discus:as a merger}.}
The spectra extracted from these regions are shown in the right column of Fig.~\ref{fig:se_zoomin}, and the measured properties are listed in Table.~\ref{tab:lines}.
\red{The \#Main region shows a bimodal distribution of the line profile characteristic of a rotating structure.
The \#Tidal region contains $\sim30$ per cent of the molecular gas in the SE galaxy and is globally blueshifted, which significantly contributes to the blueshifted SE - 1 in Cycle 6.}

\subsection{Molecular gas mass}
\label{sec: gas mass}
We calculated the integrated source brightness temperature from CO(6-5) line intensity using the following equation \citep{2013ARA&A..51..105C}:
\begin{equation}
    L_\text{CO(6-5)}^\prime=3.25\times10^7\times \frac{S_\text{CO(6-5)}\Delta\nu D_L^2}{(1+z)^3\nu_{\rm obs}^2}\ \mathrm{K\ km\ s^{-1}\ pc^2},
\end{equation}
where $\tnr{S}{CO(6-5)}\Delta\nu$ is the integrated intensity of the CO(6-5) line in $\mathrm{Jy\cdot km\ s^{-1}}$, $D_L$ is the luminosity distance in Mpc, and $\tnr{\nu}{obs}$ is the observed frequency of the CO(6-5) line in GHz.
To estimate the molecular gas mass, $L_\text{CO(6-5)}^\prime$ has to be converted to $\tnr{L}{CO(1-0)}^\prime$ via the line intensity ratio $\tnr{r}{65/10}=\tnr{S}{CO(6-5)}\Delta\tnr{\nu}{CO(6-5)}/\tnr{S}{CO(1-0)}\Delta\tnr{\nu}{CO(1-0)}$.
And the molecular gas can then be calculated using \citep{2013ARA&A..51..105C}:
\begin{equation}
    M_\mathrm{H_2}=\alpha_\text{CO}\times L_\text{CO(1-0)}^\prime\ \mathrm{\msun},
\end{equation}
where $\alpha_\text{CO}$ is the CO-to-H$_2$ conversion factor.
An $\tnr{\alpha}{CO}\sim0.8\ \msun\ \mathrm{(K\ \kms\ pc^2)}^{-1}$ \citep{1998ApJ...507..615D} found in the starburst nuclei of ULIRGs on scales $<1$ kpc is adopted as a conservative estimation.

The CO(1-0) line emission was observed by the Australia Telescope Compact Array (ATCA) with a beam size of $4.0\arcsec\times1.3\arcsec$, incapable of resolving two galaxies \citep{2015MNRAS.451.1025E}.
By tapering CO(6-5) observation from Cycle 2 to the same resolution as the CO(1-0) data, \citet{2015A&A...584A..99E} compared CO(6-5) and CO(1-0) spectra extracted from the same region covering the entire Dragonfly galaxy and found that the bulk molecular gas component has a line intensity ratio $\tnr{r}{65/10}\sim13$.
Scaling the CO(1-0) spectrum by 13 and subtracting it from the CO(6-5) spectrum, \citet{2015A&A...584A..99E} found large CO(6-5) residuals corresponding to the high-excitation gas at the blueshifted side with $v\leq-200\ \kms$.
Including $2\sigma$ measurement uncertainties as an upper limit for $\tnr{I}{CO(1-0)}$, \citet{2015A&A...584A..99E} loosely constrained $17\leq\tnr{r}{65/10(blue)}\leq36$ for the blueshifted, high-excitation components.
We adopt this high ratio for molecular gas mass estimation of the blueshifted components and the corresponding $\tnr{M}{H_2}$ are listed in Table.~\ref{tab:lines}.

\begin{table*}
\caption{CIGALE user-specified components for SED fitting described in \S\ref{sec:sed fitting}}
\label{tab:cigale parameters}
\begin{tabular}{c|c}
\hline
Parameter & Values \\
\hline
\multicolumn{2}{c}{Delayed star formation history with additional burst: $\texttt{delayedSFH}$ \citep{2015AA...576A..10C} } \\
e-folding time [Myr]: $\tau_\mathrm{main}$ & 1000, 2000, 5000, 8000 \\
Age of the main stellar population [Myr]: $t_\mathrm{main}$ & 100, 500, 1000, 1500, 2000, 4000 \\
e-folding time of the starburst [Myr]: $\tau_\mathrm{burst}$ & 10, 20, 50, 100, 200 \\
Age of the starburst [Myr]: $t_\mathrm{main}$ & 10, 20, 50, 100 \\
Mass fraction attributed to starburst & 0.1, 0.3, 0.5 \\
\hline
\multicolumn{2}{c}{Single-age stellar population (SSP): $\texttt{BC03}$ \citep{2003MNRAS.344.1000B} }\\
Initial mass function & \citet{2003PASP..115..763C} \\
Metallicity: $Z$ & 0.02 (fixed) \\
\hline
\multicolumn{2}{c}{Nebular Emission \citep{2011MNRAS.415.2920I} }\\
Ionization parameter: $\log{U}$& -3.0 (fixed) \\
\hline
\multicolumn{2}{c}{Dust attenuation: $\texttt{dustatt\_modified\_starburst}$ \citep{2000ApJ...533..682C} }\\
Interstellar reddening of the nebular lines: $E(B-V)_{\mathrm{lines}}$ & 0.5, 1, 1.5 \\
Fraction of continuum to line color excess: $E(B-V)_\mathrm{factor}$ & 0.44 (fixed) \\
\hline
\multicolumn{2}{c}{Dust emission: $\texttt{dl2014}$ \citep{2007ApJ...663..866D,2014ApJ...780..172D} }\\
Mass fraction attributed to PAH emission: $q_\mathrm{PAH}$ & 0.47, 1.12, 1.77, 2.50  \\
Minimum radiation field: $U_\text{min}$ & 10, 30, 50  \\
Power-law slope relates dust mass and radiation field intensity: $\alpha$ & 2, 3 \\
Fraction illuminated from $U_\text{min}$ to $U_\text{max}$: $\gamma$ & 0.02 (fixed) \\
\hline
\multicolumn{2}{c}{Clumpy two-phase torus model: $\texttt{SKIRTOR2016}$ \citep{2012MNRAS.420.2756S,2016MNRAS.458.2288S}} \\
Average edge-on optical depth at $9.7\ \micron$: $\tau$ & 3, 7, 11 \\
The angle between the disc plane and dust torus [${}^\circ$]: $\theta$ & 60, 70, 80 \\
The inclination angle relative to the LOS [${}^\circ$]: $i$ & 20, 30, 40 \\ 
AGN fraction measured as the IR luminosity of AGN to the total value: $f_\text{AGN}$ & 0.1, 0.3, 0.5 \\ 
\hline
\end{tabular}
\end{table*}

\section{SED Fitting}\label{sec:sed fitting}
\subsection{Optical-to-FIR SED}
The electromagnetic radiation emitted by galaxies at multiwavelength provides proxies to investigate the formation and evolution of the galaxy.
Fitting the observed spectral energy distribution (SED) with a combination of various templates allows us to disentangle the galaxy emission complexity and compute both host galaxy and AGN physical properties.
We performed SED fittings using CIGALE (Code Investigating GALaxy Emission; \citealt{2005MNRAS.360.1413B,2009A&A...507.1793N,2019A&A...622A.103B}).
Based on the new photometric data from optical-to-near IR (NIR) collected from the Dark Energy Camera (DECam) constructed for the Dark Energy Survey (DES; \citealt{2018ApJS..239...18A,2021ApJS..255...20A}) in addition to the existing IR photometry with \textit{Spitzer} and \textit{Herschel}, we re-examine the physical properties including SFR and $M_\star$ and compare them with those reported in the literature \citep{2014AA...566A..53D,2010ApJ...725...36D,2019AA...621A..27F}.
All photometric data used for the SED fitting are summarized in Table.~\ref{tab:photometry}.
We note that, since there are no spatially resolved optical-to-far IR (FIR) data available, the SED fitting thus can only compute the physical properties of the entire system, including NW and SE galaxies.

Recent studies of AGN host galaxies and LIRG populations through SED fitting favor a star formation history (SFH) with a recent burst \citep{2021A&A...649L..11T,2022ApJ...938..152D, 2023arXiv230200530G}.
This is naturally expected for galaxy mergers, especially late-stage mergers that have small bulge separations.
In these systems, the gas inflow -- attributed to angular momentum removal by the gravitational torque -- towards the galaxy center may enhance the SFR within this central region \citep{2015MNRAS.448.1107M,2021MNRAS.503.3113M,2022MNRAS.515.3406M, 2022ApJ...940....4S}.
Therefore, we assumed a delayed SFH model with an additional burst for the host galaxy \citep{2015AA...576A..10C}.
In addition to the e-folding time $(\tau_\text{main})$ and age $(t_\text{main})$ of the main stellar population, $\tau_\text{burst}$, $t_\text{burst}$, and a mass fraction of the late burst population $(f_\text{burst})$ can be parameterized.

We adopted \citet{2003MNRAS.344.1000B} stellar population synthesis library to model the stellar emission with a solar metallicity 0.02, Chabrier initial mass function \citep{2003PASP..115..763C}, and a standard model for nebular emission \citep{2011MNRAS.415.2920I}.
Although AGNs would enrich their environments, resulting in higher chemical abundance, we consider such an enrichment negligible in this HyLIRG \citep{2013MNRAS.431..793Z, 2015MNRAS.452L..59T}.
The dust emission was modeled adopting a \texttt{dl2014} template \citep{2007ApJ...663..866D,2014ApJ...780..172D}.
This template includes a parameter $q_\mathrm{PAH}$ which optimizes the mass fraction of the polycyclic aromatic hydrocarbon (PAH) emission that dominates the MIR emission, a radiation field $U_{\rm min}$ that models the diffuse dust emission, and a power-law distribution of the dust corresponding to the star-forming regions with index $\alpha$.
The $\alpha$ is defined in a way such that $d\tnr{M}{d}(U)/dU\propto U^{-\alpha}$, where $\tnr{M}{d}$ is the dust mass and U is the radiation field intensity.
The initial estimations of these parameters are referred from recent studies of AGN host galaxies \citep{2023arXiv230103613Y,2021A&A...654A..93B}.

UV and optical emissions from the AGN will be absorbed and then re-emitted at longer wavelengths owing to the existence of the obscuring structure, i.e., the dusty torus. 
To simulate this reddening effect, we adopted SKIRTOR, a clumpy two-phase torus model derived from a modern radiative-transfer method, assuming that the dusty torus is made of dusty clumps rather than a smooth structure \citep{2012MNRAS.420.2756S,2016MNRAS.458.2288S}. 
This model depends on several parameters including the average edge-on optical depth at 9.7 $\micron$, the half-opening angle of the dust-free cone, inclination (\textit{i}, $0^\circ$ for face-on and $90^\circ$ for edge-on), and the AGN fraction defined as the fraction of AGN IR luminosity to total (AGN$+$host) IR luminosity.

The models described above and the values of corresponding parameters are summarized in Table.~\ref{tab:cigale parameters}.

\subsection{Radio SED}
In the radio regime, CIGALE models the flux densities attributed to different mechanisms, including synchrotron radiation from AGN and star formation and nebular continuum emission that involves mainly free-free emission \citep{2019A&A...622A.103B}.
There are four parameters controlling the model of the synchrotron radiation: the FIR-to-radio correlation parameter $\tnr{q}{IR}$, radio loudness $\tnr{R}{AGN}$ defined as the ratio of $\tnr{L}{5GHz}/\tnr{L}{2500\AA}$ where $\tnr{L}{2500\AA}$ reflects the AGN disc luminosity, and the spectral index of star formation- ($\tnr{\alpha}{SF}$) and AGN-related synchrotron radiation $\tnr{\alpha}{AGN}$.

In normal star-forming galaxies (SFGs), the synchrotron radiation is primarily produced by the electrons accelerated in supernova remnants.
In the Milky Way and many other SFGs, the overall slope of the observed radio SED has been known to have a mean value of $\tnr{\alpha}{SF}\approx-0.75$ at $\nu\approx1$ GHz in a negative convention ($S\propto\nu^\alpha$) \citep{2016era..book.....C,2018A&A...611A..55K}.
Hence, we fix $\tnr{\alpha}{SF}$ to -0.75 for the synchrotron radiation relevant to the star formation.

In the Dragonfly galaxy, the bulk of the synchrotron radiation originates from the SE hotspot and its associated diffuse radio lobe and can be significantly enhanced (see \S\ref{sec: jet luminosity}).
Therefore, the observed radio emission no longer traces the intrinsic AGN activities but the physical conditions in situ.
Thanks to VLA 44 GHz observations, the radio core associated with the RLAGN in the NW galaxy has been identified (Fig.~\ref{fig:imaging}).
Such a core component generally has a spectral slope flatter than those of radio hotspots and lobes because of synchrotron self-absorption and/or free-free absorption arising from the dense environment.
We adopt a spectral index of $\alpha_\text{AGN}=0$, which is a commonly observed value in flat-spectrum radio quasars that are core-dominated populations \citep{2009MNRAS.397.1713C}.
$\tnr{R}{AGN}$ and $\tnr{q}{IR}$ are fixed at 10 and 2.3, respectively.

So far, the SED fittings for powerful radio galaxies primarily make use of $\tnr{S}{1.4GHz}$ without spatially decomposing the emission source and considering enhancements on the flux densities.
The applications of $\tnr{q}{IR}$ and $\tnr{R}{AGN}$ on the radio fitting are highly motivated in this manner.
Therefore, a consideration of only the radio core, excluding radio hotspots and lobes, for the Dragonfly galaxy may lead to a systematic bias.
We then perform a fitting using 1.4 GHz data instead of the 44 GHz data assuming a spectral index of $\alpha_\text{AGN}=1.0$ as a comparison.
In this fitting, we fixed $\tnr{R}{AGN}=1000$ and $\tnr{q}{IR}=0$.

\begin{table*}
\caption{Host galaxy and AGN physical properties computed using CIGALE}
\label{tab:fitting result}
\begin{tabular}{lllllll}
\hline
Parameter & 44 GHz${}^a$ & 1.4 GHz${}^b$ & No radio & \citet{2014AA...566A..53D} & \citet{2019AA...621A..27F} & \citet{2010ApJ...725...36D} \\
\hline
$M_\star\ [\times10^{10}\ \msun]$ & $9.2\pm2$ & $9.2\pm4$ & $11\pm3$ & - & - & $58$ \\
Recent burst fraction & $0.34\pm0.15$ & $0.46\pm0.16$ & $0.27\pm0.16$ & - & - & - \\
SFR $[\msun\ \per{yr}{-1}]$ & $2700\pm300$ & $2100\pm100$ & $2300\pm200$ & $3100\pm200$ & $1900$ & - \\
SFR$_\text{10Myr}$ $[\msun\ \per{yr}{-1}]$ & $3100\pm400$ & $2200\pm200$ & $2500\pm300$ & - & - & - \\
SFR$_\text{100Myr}$ $[\msun\ \per{yr}{-1}]$ & $600\pm200$ & $900\pm300$ & $900\pm200$ & - & - & - \\
Main age [Myr] & $930\pm710$ & $570\pm650$ & $650\pm670$ & - & - & - \\
Burst age [Myr] & $13\pm6$ & $36\pm26$ & $28\pm25$ & - & - & - \\
$L\mathrm{_{AGN,bol}\ [\times10^{46}\ \ergs]}$ & $0.9\pm0.1$ & $2.9\pm0.1$ & $3\pm0.3$ & $1.7\pm0.5$ & $3.7\pm0.5$ & - \\
AGN fraction & $\sim0.1$ & $\sim0.3$ & $\sim0.3$ & $\sim0.2$ & $\sim0.4$ & - \\
\hline
\end{tabular}
\begin{tablenotes}
{\item \raggedright  ${}^a$The spatially resolved 44 GHz flux density of the radio core that overlaps with the NW galaxy is used for the fitting.\\
${}^b$The unresolved total 1.4 GHz flux density, including the radio core, hotspots, and lobes is used for the fitting.}
\end{tablenotes}
\end{table*}

\subsection{Fitting Results}\label{SED results}
\begin{figure}
\begin{center}
\includegraphics[width=0.48\textwidth]{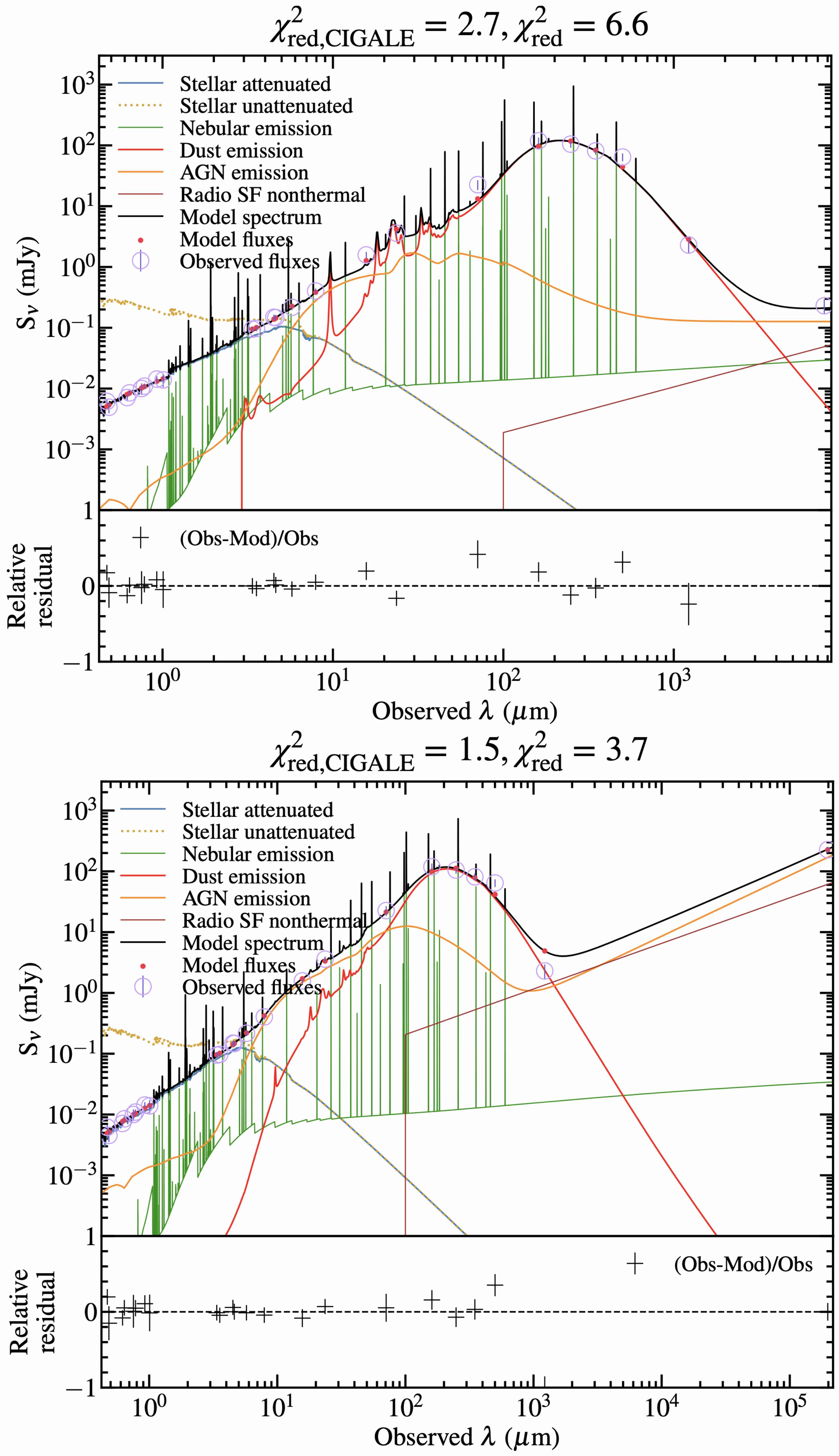}
\caption{\label{fig:sed_radio}
The best fitting models of UV-to-radio SED for the Dragonfly galaxy.
The upper figure shows the fitting up to 44 GHz of the radio core. 
The lower one is up to 1.4 GHz without the 44 GHz flux density.
In each figure, the upper panel shows the observed and modeled flux densities against the observed wavelength and the lower panel shows the residuals of the model in each corresponding band.
The reduced chi-square defined by CIGALE is $\chi^2_{\rm red,CIGALE}=\chi^2/(\rm data\ points-1)$, where the degree of freedom is not involved.
We re-calculated $\chi^2_{\rm red}$ considering the number of fitted free parameters.
}
\end{center}
\end{figure}

\red{We have performed three fits to the SED: firstly based on $\tnr{S}{1.4 GHz}$ of the total radio emission, then based on $\tnr{S}{44 GHz,core}$ of the radio core, and finally without radio data.}
We summarize the best-fitting results in Table.~\ref{tab:fitting result} and plot the modeled SEDs in Fig.~\ref{fig:sed_radio}.
The stellar mass of NW plus SE galaxies is $M_{\rm \star,NW+SE}\sim(9-11)\times10^{10}\rm\ M_\odot$ with an instantaneous SFR of $\sim2000-3000\ \rm M_\odot\ yr^{-1}$.
For all fittings, the instantaneous SFR is close to the SFR averaged over 10 Myr but beyond that averaged over 100 Myr, suggesting recent starbursts.
We performed a test without an additional starburst by fixing the starburst mass fraction to 0.
Different from the fiducial fitting with the recent burst, the test fitting includes only a single population that forms within 100 Myr, still suggesting recent star-forming activities.

\subsubsection{Comparisons with literature}
Using \textit{Spitzer} and \textit{Hershcel} data (3-500 $\micron$), \citealt{2014AA...566A..53D} decomposed the IR luminosity into AGN and starburst components by simultaneously calculating AGN and host galaxies contributions to the SED.
They found $L\mathrm{^{IR}_{SB}=1.72\times10^{13}\ \lsun}$ attributed to star-forming activities and estimated $\mathrm{SFR\sim3000\ \sfr}$ based on the SFR and IR luminosity relation \citep{1998ARA&A..36..189K}.
Using a non-starburst SED template and combining low-frequency VLA observations at 1.4, 4.7, and 8.2 GHz, \citet{2019AA...621A..27F} found $\mathrm{SFR\sim1900\ \sfr}$. 
Our SED fitting analysis involving an SFH with an additional starburst finds $\mathrm{SFR\sim2100-2700\ M_\odot\ yr^{-1}}$ that agrees with \citet{2014AA...566A..53D} and slightly higher than \citet{2019AA...621A..27F}.
The stellar mass of our fitting is merely one-fifth the value of $M_\star\sim5\times10^{11}\ \msun$ reported in \citet{2010ApJ...725...36D}.
Such a difference could be explained by the choice of elliptical galaxy templates instead of a star-forming one in the SED fitting performed by \citet{2010ApJ...725...36D}, the lack of IR data at $\lambda\geq70\ \micron$, and the lack of optical photometry. 
An elliptical galaxy SED dominated by old stellar populations is a reasonable choice for low-redshift populations since powerful radio galaxies in the nearby Universe are found to preferentially reside in massive, early-type galaxies (ETGs), typical of the final evolution stage of an AGN \citep{2008ApJS..175..356H,2020NatAs...4..282S}.
However, RLAGNs at high redshifts are often hosted by mergers that are often accompanied by intense star-forming activities, thus the choice of SED templates of ETGs is not always proper for HzRGs.

The AGN has an IR contribution of $\sim0.1$ to the total IR luminosity given by our fitting including 44 GHz data, significantly smaller than the value of $\sim0.4$ given by \citet{2019AA...621A..27F}.
Our fitting including the observed 1.4 GHz flux density, as well as the one without radio data, returns a similar AGN fraction of 0.3 compared with $\sim0.2$ given by \citet{2014AA...566A..53D} and $\sim0.4$ given by \citet{2019AA...621A..27F}.
These results are naturally expected since when the radio data is included in our fitting, the AGN disc luminosity depends on the normalization of the luminosity at $2500\ \r{A}$ using $\tnr{R}{AGN}$.
The 1.4 GHz data can obviously lead to a higher luminosity at $2500\ \r{A}$ than 44 GHz data does because of a much larger radio loudness.
The AGN component dominates over the dust component up to $\sim80\ \micron$ in the 1.4 GHz case and merely up to $\sim20\ \micron$ in the 44 GHz case.
Accordingly, the AGN contributes more to the total IR luminosity, resulting in a larger $\tnr{f}{AGN}$.

\subsubsection{Caveats to fitting results}\label{sec: SED caveats}
Since there is no spatially resolved optical-to-FIR data available for the SED fitting, the reported physical properties regarding the AGN may not be the proxy to the RLAGN that resides in the NW galaxy solely.
If the SE galaxy also has an AGN, though we find no signature yet, the AGN component presented here is the sum of the NW AGN and the potential SE AGN.

The fitting of synchrotron emission does not include detailed physical processes -- how jets interact with the medium in situ -- but depends on the empirical $\tnr{q}{IR}$ and $\tnr{R}{AGN}$ for radio galaxies with lobes and/or hotspots. 
And the finally derived $\tnr{L}{AGN,bol}$ will be regulated by the $\tnr{R}{AGN}$ when radio data is included.
Therefore, a fitting based on the radio core but neglecting radiation from radio hotspots/lobes requires values of $\tnr{q}{IR}$ and $\tnr{R}{AGN}$ for weaker AGNs, putting a lower limit on the $\tnr{L}{AGN,bol}$ in our target.
\red{On the other hand, the imbalanced flux densities between SE and NW hotspots at 1.4 GHz are suggestive of possibly enhanced synchrotron radiation due to the environmental difference or the Doppler boosting effect (see Paper I, \citealt{2023MNRAS.522.6123Z} and  \S\ref{sec: jet luminosity}).}
Then, the fitting requires values of $\tnr{q}{IR}$ and $\tnr{R}{AGN}$ for more powerful AGNs, which may result in overestimated $\tnr{L}{AGN,bol}$.
After removing the radio data, we find $\tnr{f}{AGN}$ and $\tnr{L}{AGN,bol}$ consistent with the results based on 1.4 GHz.
This may suggest that the application of $\tnr{q}{IR}$ and $\tnr{R}{AGN}$ is reliable even when dealing with powerful radio galaxies with biased flux densities.
However, we still call for caution when trying to retrieve $\tnr{L}{AGN,bol}$ for these HzRGs.

In conclusion, the Dragonfly galaxy is a massive, starburst one with $M_\star\sim(0.9-1.1)\times10^{11}\rm\ \msun$ and an SFR of $\rm\sim2000-3000\ M_\odot\ yr^{-1}$.
The RLAGN reaches a QSO-level luminosity of $\tnr{L}{AGN,bol}\sim(0.9-3)\times10^{46}\ \ergs$.
We can take the fitting using 44 GHz as the lower limit for AGN properties, which sufficiently suggests that the RLAGN is radiatively efficient ($\tnr{\lambda}{Edd}>3\times10^{-2}$; see \S\ref{sec: bh mass} for detailed discussions).
In \S\ref{sec: agn activities} and \S\ref{sec: AGN feedback}, we will further discuss the AGN activities by investigating the Eddington ratio and outflow properties.

\section{Gas Kinematics}
\label{sec: kinematics}
\begin{figure*}
\begin{center}
\includegraphics[width=\textwidth]{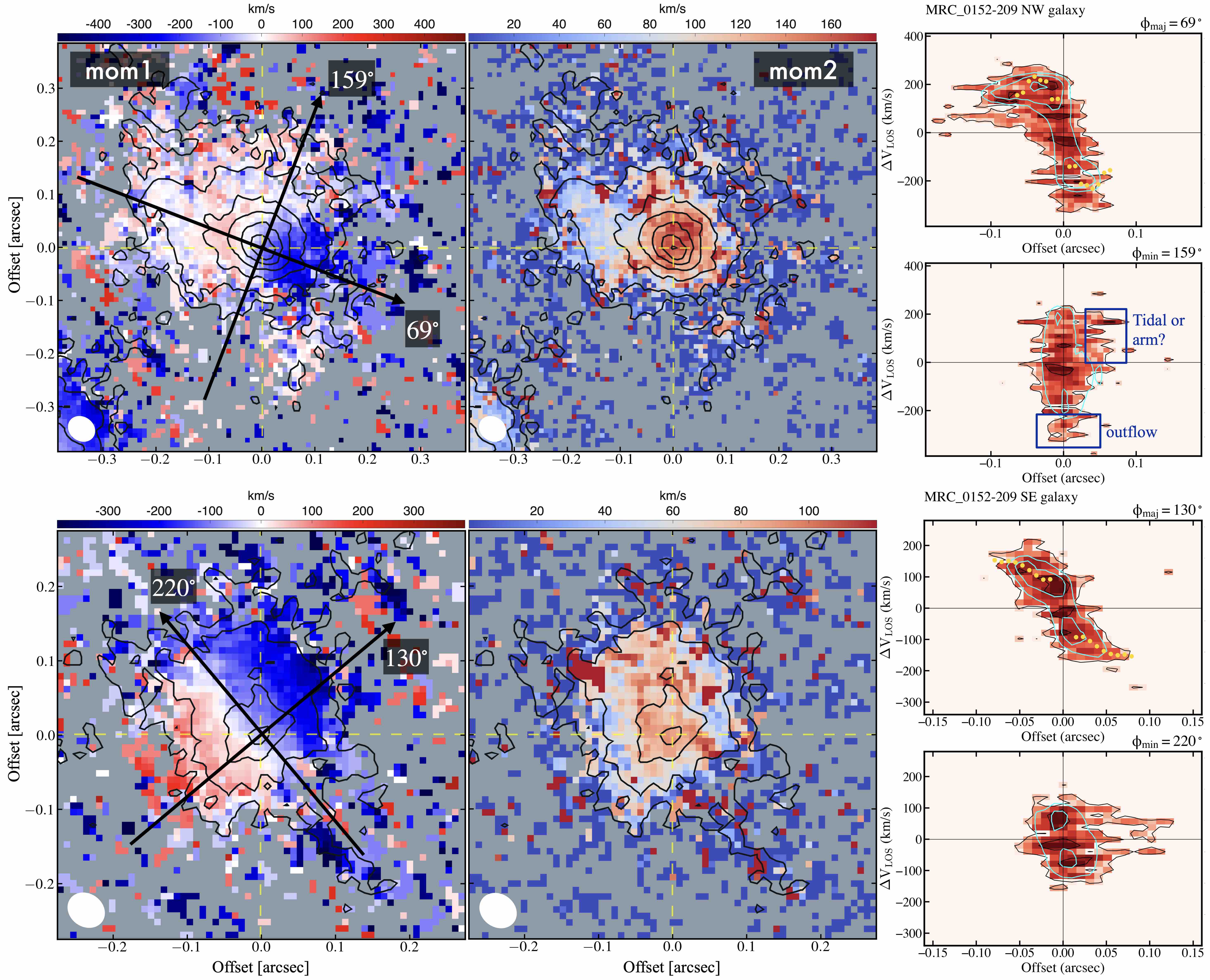}
\caption{\label{fig:barolo}
\red{From left to right, the panels show the velocity field (mom1), the velocity dispersion (mom2) maps generated from the combined C46 dataset by excluding pixels under $3\sigma$, and the position velocity diagrams (PVDs) extracted from the observed and modeled data cube.
The upper row shows the results from the NW galaxy and the lower row for the SE galaxy.}
In PVDs, the color code (with black contours) indicates the observational data, the cyan contours indicate the modeled data, and the yellow dots indicate the modeled rotational velocity.
$\Delta\tnr{V}{LOS}$ is the observed velocity field $\tnr{V}{LOS}$ corrected for systemic velocity ($\Delta\tnr{V}{LOS}=\pm||\tnr{V}{LOS}|-|\tnr{V}{sys}||$).
The black arrows in mom1 maps indicate the slits of major and minor axes used to extract the PVD.
The slits here only indicate the extraction direction and the slit width is approximately the beam size.
}
\end{center}
\end{figure*}

\subsection{Modeling with $^\text{3D}$Barolo}
We investigated the molecular gas kinematics by fitting a kinematic tilted-ring model to the CO(6-5) line emission data cube using $^\text{3D}$Barolo \citep{2015MNRAS.451.3021D}.
$^\text{3D}$Barolo assumes that the rotating gaseous disc of a galaxy is made up of a series of concentric rings, where each ring is described by geometrical and kinematical parameters, including spatial center ($x_0$, $y_0$), systemic velocity ($V_\text{sys}$), position angle (PA), inclination angle (\textit{i}), scale height of the disc $(z_0)$, rotation velocity ($V_\text{rot}$), and velocity dispersion ($\sigma_\text{V}$).

The modelings are based on the C46 dataset.
We adopted 8 rings for the NW galaxy and 9 rings for the SE galaxy.
We chose a radial separation of 1/5 along the minor axis of the beam \citep{2022A&A...658A.155R}.
We tested modelings with larger separations up to $\rm beam_{minor}/2$ and found similar results.
The outermost ring of each galaxy was limited to be smaller than the aperture size that includes $3\sigma$ contours in the mom0 map to exclude possible sub-structures such as tidal tails/spiral arms.
None the less, as we will see in \S\ref{Kine: NW galaxy} and Fig.~\ref{fig:barolo}, these features still contaminate the velocity field on the redshifted side of the NW galaxy.
We set the scale height as 0 adopting a thin disc approximation where the effect of any vertical structure is neglected \citep{2022A&A...658A.155R,2022MNRAS.516.4066L}. 
We fixed the kinematic centers of both galaxies to the coordinates of 1.2 mm dust continuum peak positions.
These coordinates were uniform such that the rings remained concentric.
The rotational velocity, velocity dispersion, systemic velocity, position angle, and inclination angle were left as free parameters.
We adopted a \texttt{twostage} fitting strategy such that the algorithm will perform a first fitting stage to estimate all free parameters.
Then, after regularizing the free parameters, the algorithm proceeds to a second stage where the radial profiles of inclination and position angles follow a Bezier function and kinematical parameters ($V_\text{rot}$ and $\sigma_\text{V}$) are allowed for free parameters \citep{2015MNRAS.451.3021D}.

In the upper row of Fig.~\ref{fig:barolo}, we show the velocity field (mom1) and velocity dispersion (mom2) maps directly generated from  the observed CO(6-5) data cubes by excluding pixels under $3\sigma$.
\red{The noise level is calculated by averaging the standard deviation of each channel of the data cube.
The position-velocity diagrams (PVDs) are shown in the lower row of Fig.~\ref{fig:barolo} for NW and SE galaxies, respectively.
The PVDs are extracted along the kinematic major and minor axes indicated by arrows in the mom1 and mom2 maps with a slit width the same size as the FWHM of the synthesized beam along the major axis, which are automatically produced by running $^{\rm 3D}$Barolo.
The width of the region used for the extraction is the same as the region shown in Fig.~\ref{fig:3dbarolo model nw} and \ref{fig:3dbarolo model se}.} 

\subsection{SE Galaxy}\label{Kine: SE galaxy}
The PVDs of the SE galaxy clearly show ordered rotation along the major axis.
A clear velocity gradient, as well as the zero velocity region, can be seen in the main structure in the mom1 map generated from Cycle 6 observations (see Fig.~\ref{fig:cycle6 moments}), which reveals a PA of $132^\circ$.
This value is consistent with $\rm PA=130^\circ$ retrieved by $^\text{3D}$Barolo, suggesting that the rotating structure in the PVD corresponds to the structure in \#Main.
This is also supported by that the main structure has a bimodal line profile characteristic of rotation (Fig.~\ref{fig:se_zoomin}).
In addition to these rotational molecular gas, the PVD along the minor axis shows non-circular motions with offset $>0.05\arcsec$.
\red{The component with $\tnr{V}{LOS}\sim-100\ \kms$ corresponds to the \#Tidal but is partially smeared because of the larger beam size of C46.
Because of the tidal effect, which may give rise to the tidal bridge between the two galaxies (see \S\ref{discus:as a merger}), the filamentary structure in \#Tidal cannot be simply treated as a part of rotation subject to the gravitational force.}

The modeling returns an inclination angle of $28^\circ$.
\red{This value varies by only $1^\circ$ from the innermost to outermost rings despite the gas being distorted by the tidal force.
With the inclination angle, we can calculate the ratio of the rotational velocity ($\tnr{V}{rot}=\tnr{V}{LOS}/\sin i$) to velocity dispersion for a series of concentric rings returned by $^{\rm 3D}$Barolo.}
These rings have a minimum ratio of $\tnr{V}{rot}/\tnr{\sigma}{V}\geq5$, indicative of a rotation-dominated disc \citep{2017ApJ...843...46S}.

\subsection{NW Galaxy}\label{Kine: NW galaxy}
The high-resolution observations reveal non-circular and tangential motions that are diluted because of the beam-smearing effect in Cycle 2.
These non-circular motions include the massive ($\tnr{M}{H_2}\sim2\times10^9\ \msun$), lopsided molecular outflow (see \S\ref{sec: AGN feedback}), which corresponds to $\rm NW-1$ in Fig.~\ref{fig:imaging}.
This outflow component is clearly reflected in the NW galaxy PVDs with $\rm V_{LOS}\leq-300\ \kms$ (Fig.~\ref{fig:barolo}).
The diffuse molecular gas in the northeast part (see C46 moment maps in Fig.~\ref{fig:barolo} and Off-2 in Fig.~2 of \citealt{2023ApJ...951...73L}) is another origin of non-circular motions and is labeled in the PVD along the minor axis.
This additional CO(6-5) line emission has $-100\lesssim\tnr{V}{LOS}\lesssim0\ \kms$ with an offset $<-0.05\arcsec$ as clearly shown in the C46 PVDs along the major axis.
In the current stage, lack of observations primarily tracing cold molecular gas, we cannot confirm whether this feature originates from the spiral arm, thus belonging to part of the rotation, or stems from tidal effects.

In the C46 mom1 map, as well as indicated by the green curve in the observational data constructed by $^\text{3D}$Barolo ("VELOCITY" in Fig.~\ref{fig:3dbarolo model nw}), the projected zero-velocity region is curved.
This curvature feature has been reproduced in the simulated rotating disc plus jet-driven outflows, showing that the kinematic centers could be curved rather than being a straight line \citep{2022MNRAS.516..766M}.
Consequently, the kinematic major axis and inclination angle of the CO disc cannot be unambiguously defined, as we will discuss below.

\red{The non-circular motions make it more difficult to model the rotational velocity field.
Depending on the initial guesses of position and inclination angles, the modeled PA ranges within $[69^\circ, 82^\circ]$, and the inclination angle ranges within $[27^\circ, 55^\circ]$.
By exploring the PA$-$inclination parameter space using the $\texttt{SPACEPAR}$ task of $^\text{3D}$Barolo, we found that the PA and inclination angles have degenerated.}
The main kinematic parameter influenced by this loosely constrained parameter space is the rotation velocity $\tnr{V}{rot}$ as it is calculated by $\tnr{V}{rot}=\tnr{V}{LOS}/\sin{i}$, where $\tnr{V}{LOS}$ is the observed velocity.
We then directly estimated the inclination angle using \citep{2014RvMP...86...47C}
\begin{equation}
    i=\cos^{-1}\sqrt{\frac{(b/a)^2-q_0^2}{1-q_0^2}},
\end{equation}
where $a$ is the semi-major axis, $b$ is the semi-minor axis, and $q_0$ is the axial ratio of a galaxy viewed edge-on.
Adopting $q_0=0.13$ for late-type galaxies \citep{2012MNRAS.425.2741H}, the inclination is $24^\circ$ based on Cycle 6 observations where the NW galaxy is spatially resolved.
Hence, although we cannot determine the precise geometry of the NW galaxy, an $i\sim27^\circ$ could be used to set an upper limit for $\tnr{V}{rot}$, which is $\sim520\ \kms$.

The modeled rings have a minimum $\tnr{V}{rot}/\tnr{\sigma}{V}$ of $\sim3$ and a maximum of $\sim50$, suggesting that the NW galaxy is rotation-dominated as well.
Although this ratio is highly varied because of the large uncertainties arising from the non-circular motions, in low-resolution Cycle 2 observations where the most non-circular motions are diluted, the velocity field map of the NW galaxy shows distinct blueshifted and redshifted regions, indicating a rotating disc \citep{2015A&A...584A..99E,2023ApJ...951...73L}.

\subsection{Dynamical Mass}\label{sec: dynamical mass}
As argued by \citet{2015A&A...584A..99E}, the two galaxies should have low inclinations ($10-20$ degrees) such that the dynamical mass is comparable with the high stellar mass $M_\star\sim5.8\times10^{11}\ \mathrm{\msun}$ computed by \citet{2010ApJ...725...36D}.
However, based on our SED fitting results, the stellar mass is $M_\star\sim(0.9-1.1)\times10^{11}\ \mathrm{\msun}$, which differs from the literature value by a factor of 5 but should be a more robust estimate (see \S\ref{SED results}).
In this case, neither the ALMA Cycle 2 observation has significantly missed the extent of rotating discs nor an extremely low inclination is required for the dynamical mass to match the stellar mass.
The inclination angle estimated from modeling gas kinematics using $^\text{3D}$Barolo is $28^\circ$ for the SE galaxy and $27^\circ$, which is likely to be the lower limit (see \S\ref{Kine: NW galaxy}), for the NW galaxy.
We then estimate the dynamical mass of each galaxy by \citep{2014RvMP...86...47C}
\begin{equation}
M_\text{dyn}=2.33\times10^5RV_\text{obs}^2/\sin^2i\ [\mathrm{\msun}],
\end{equation}
where $R$ is the radius along the major axis, $V_\text{obs}$ is the observed rotation velocity, and $i$ is the inclination angle.
\red{$R$ is determined by fitting a 2D Gaussian component to the mom0 map of Cycle 6 observations and being deconvolved from the beam size.}
For the NW galaxy with $R=0.8$ kpc and $i=27^\circ\pm2^\circ$, the observed rotation velocity is $V_\text{obs}=236^{+23}_{-32}\ \kms$, which gives a dynamical mass of $M_\text{dyn,NW}=5.0^{+1.1}_{-1.2}\times10^{10}\ \mathrm{\msun}$.
And for the SE galaxy with $R=0.8$ kpc, $V_\text{obs}=155^{+8}_{-10}\ \kms$, and $i=28^\circ\pm2^\circ$, the dynamical mass is $M_\text{dyn,SE}=2.0^{+0.3}_{-0.3}\times10^{10}\ \mathrm{\msun}$.

The dynamical mass can be compared with other mass estimates \citep[e.g.,][]{2022A&A...665A.107T}.
Without considering the BH mass, the NW galaxy has a stellar mass of $\tnr{M}{\star,NW}=\tnr{M}{dyn,NW}-\tnr{M}{H_2,NW}\sim3\times10^{10}\ \msun$ within a 0.8 kpc radius.
This rough estimate likely rejects a large inclination angle $(i\gtrsim40^\circ)$.
Otherwise, the dynamical mass will be even smaller than the molecular gas mass.
For the SE galaxy, $\tnr{M}{\star,NW}\sim\tnr{M}{dyn,SE}-\tnr{M}{H_2,SE}\sim1\times10^{10}\ \msun$ within a 0.8 kpc radius.
Within a 0.8 kpc radius, $\tnr{M}{\star,NW+SE}$ is expected to be $\sim4\times10^{10}\ \msun$, significantly smaller than the value $9.2\times10^{10}\ \msun$ returned by the SED fitting.
This suggests the existence of extended stellar components out of the CO discs, which can be seen from Fig.~\ref{fig:hst imaging} that the stellar continuum extends well beyond the CO(6-5).
A size comparison between stellar and molecular gas components in local LIRGs has been done by \citet{2022A&A...664A..60B}, finding out that the stellar size traced by ionized gas is typically three times that of the molecular gas.

\red{\section{AGN Properties}}
\subsection{BH mass}
\label{sec: bh mass}
\red{One crucial quantity in understanding AGN evolution is the BH mass.
To provide an estimate of $\tnr{M}{BH}$ related to the RLAGN hosted by the NW galaxy,} we start with the scaling relation between central SMBH masses and the host galaxy stellar masses \citep{2013ARA&A..51..511K}.
The SED fitting returns $M_\star=(9.2\pm0.4)\times10^{10}\ \msun$ including NW, SE, and the possible companion galaxy.
Since there are no spatially resolved data allowing for an estimate of the individual galaxy, \red{we then set an upper limit of $M_{\rm\star,NW}\sim8\times10^{10}\ \msun$ for the NW galaxy such that it does not exceed the total stellar mass in this system after considering $M_{\rm\star,SE}\sim1\times10^{10}\ \msun$ for the SE galaxy within the CO disc.
The lower limit is $3\times10^{10}\ \msun$, which matches $M_{\rm \star,NW}\sim M_{\rm dyn,NW}-\tnr{M}{H_2,NW}$ estimated from the dynamical mass of $\sim5\times10^{10}\ \msun$ (see \ref{sec: dynamical mass} for details).}
We note that these assumptions do not lead to significant quantitative changes in the estimates of the Eddington ratio to be discussed in \S\ref{sec: eddington} that are directly related to the BH mass and all successive discussions because the scaling relations are intrinsically scattered.

The relation between the BH and stellar masses can be parameterized as $\log\tnr{M}{BH}=\alpha+\beta\cdot\log\left(\frac{M_\star}{10^{10}\ \msun}\right)$.
We adopt the values for $\alpha$ and $\beta$ parametrically modeled by \citet{2021MNRAS.508.3463G} for X-ray-selected AGNs up to $z=4$, which gives $M_\text{BH}\sim(0.2-10)\times10^{8}\ \msun$.

\subsection{Jet kinetic power}
\label{sec: jet luminosity}
The total kinetic power of AGN jets can be estimated from the flux densities of the radio hotspots.
We have confirmed the existence of the SE and NW hotspots in Paper I, and the NW hotspot is situated towards the northwest of the radio core, along the jet axis as shown in Fig.~\ref{fig:imaging} (see also Fig.~1 in \citealt{2023MNRAS.522.6123Z}).
The observed flux density of the SE hotspot is larger than that of the NW one by at least a factor of 6, which is argued to be a result of Doppler boosting and/or a jet-ISM interaction in the SE galaxy \citep{2023MNRAS.522.6123Z}.
Therefore, a restoration of the intrinsic flux density is required to estimate $\tnr{L}{jet}$.
In a Doppler boosted case, the intrinsic $(\tnr{S}{int})$ and observed $(\tnr{S}{obs})$ are related by $\delta^{2-\alpha}<\frac{\tnr{S}{obs}}{\tnr{S}{int}}<\delta^{3-\alpha}$, where $\alpha$ is the spectral index defined as $S_\nu\propto\nu^\alpha$, $\delta\equiv[\gamma(1-\beta\cos\theta)]^{-1}$ is the Doppler factor, and $\gamma\equiv(1-\beta^2)^{-1/2}$ is the Lorentz factor.
Following the scenario proposed in Paper I, that is, the SE hotspot has its flux density enhanced whereas the NW one dimmed, the upper limit of intrinsic flux density is then the case the NW hotspot is dimmed by a factor of four, i.e., $\delta=4$.
On the other hand, the imbalanced flux densities may be only attributed to that NW and SE jets interact with the medium of different electron densities.
In this case, we can set a lower limit for the intrinsic flux density, that is $\delta=1$.
Assuming $\alpha=-1.2$ between 1.4 and 4.7 GHz, the intrinsic flux density for a single jet at 1.4 GHz is then $\tnr{S}{int,1.4}=123$ mJy for $\delta=4$ and 31 mJy for $\delta=1$, respectively.
Adopting a relation between $\tnr{L}{jet}$ and the radio power at 1.4 GHz \citep[$\tnr{P}{1.4GHz}=4\pi D^2_{\rm L}(1+z)^{-(\alpha+1)}\tnr{S}{int,1.4}\tnr{\nu}{1.4GHz}$;][]{2010ApJ...720.1066C}:
\begin{equation}
    \tnr{L}{jet}\approx5.8\times10^{43}(\tnr{P}{1.4GHz}/10^{40})^{0.7}\ \mathrm{erg\ \per{s}{-1}},
\end{equation}
the total jet kinetic power then ranges from $2\times10^{46}\ \ergs$ for the Doppler boosted case and $4\times10^{46}\ \ergs$ for the jet-medium interaction case, respectively.

\section{Discussion}
\subsection{The Dragonfly Galaxy as a Merger}
\label{discus:as a merger}
\subsubsection{A Late-stage Merger}
\label{discus:late-stage merger}
Low-resolution ALMA Cycle 2 observations have revealed a bridge-like structure that connects NW and SE galaxies and was interpreted as tidal debris that arises from the gravitational interaction between the two rotational gaseous discs \citep{2015A&A...584A..99E}.
This speculated structure has been further investigated by \citet{2023ApJ...951...73L}, arguing that this structure is the tidal bridge with $\tnr{M}{H_2}=(3\pm1)\times10^9\ \msun$, which is often observed in major mergers in both simulations and observations \citep{2017ApJ...836...66S,2022MNRAS.509.2720S}.
Combining our kinematic modelings (see \S\ref{sec: kinematics}), these results support the consensus that the Dragonfly galaxy is a major merger constituting two likely rotating discs \citep{2015A&A...584A..99E}.

A merger will undergo a pair phase between the first and the second pericentric passage and a merging phase after the second pericentric passage and before a full coalescence \citep{2022MNRAS.515.3406M}.
During the passage, gravitational torques will exert on the gas, \citep{1996ApJ...464..641M,1996ApJ...471..115B}, leading the gas material to lose angular momentum and inflow towards the centers of the merging galaxies, fueling star formation activities.
The tidal bridge seen in Cycle 2 and the tails revealed by HST/NICMOS F160W image \citep{2001ApJS..135...63P,2015A&A...584A..99E} serve as signatures by which the Dragonfly galaxy at least can be classified as a major merger at stage 2 characterized by obvious tidal bridges and tails \citep{2016ApJ...825..128L}.
They are evident that the Dragonfly galaxy has already undergone the first pericentric passage.
Additionally, recent studies \citep[e.g,][]{2021ApJ...923...36S} of dual and offset AGNs linked to major mergers have found significantly enhanced AGN activities at bulge separations ranging from  $14-11$ kpc, attributed to the first pericentric passage, and $4-2$ kpc, attributed the second passage.
Considering also the small nuclear separation of $\sim4$ kpc and high IR luminosity $(L\mathrm{_{IR}\sim2\times10^{13}\ \lsun})$, the Dragonfly galaxy could be a late-stage merger on its way to coalescence.

The starburst activities are thus an outcome of the second pericentric passage.
This is supported by our results of SED fitting adopting an SFH with an additional recent burst, which shows that there is no significant difference between the instantaneous SFR and $\mathrm{SFR_{10Myr}}$ averaged over 10 Myr, but a comparably low $\mathrm{SFR_{100Myr}}$ averaged over 100 Myr for all three fittings (see Table.~\ref{tab:fitting result}).
In this scenario, the intense gas inflow accompanied by this passage results in a starburst that contributes to at least thirty per cent of the total stellar mass, which is reflected in the recent burst fraction $\gtrsim0.3$ in Table~\ref{tab:fitting result}.

\subsubsection{Beads-on-a-string?}
\label{sec: beads on a string}
In late-stage mergers that truly have a final coalescence, overlapped gaseous and stellar discs are commonly observed \citep{2007A&A...468...61D}.
Such overlapping of the stellar components between SE and NW galaxies traced by rest-frame UV and optical photometry can be found in HST/WFPC2 F814W and NICMOS F160W images \citep{2001ApJS..135...63P,2015A&A...584A..99E}.

From the aspect of gaseous discs, both SE and NW galaxies have their molecular contents concentrated within a sub-kpc scale region, showing no recognizable overlap other than the tidal bridge \citep{2023ApJ...951...73L}.
In Cycle 6 observations, the SE galaxy has a significant fraction ($\gtrsim30$ per cent) of the molecular gas originating from the filamentary structure in \#Tidal, which is constituted of several clumps, as shown in Fig.~\ref{fig:se_zoomin}.
In the image of the dust thermal continuum, the NW galaxy has a diffuse and extended structure elongated towards the east.
The potential tidal bridge, argued by \citet{2015A&A...584A..99E} and \citet{2023ApJ...951...73L}, links these two structures in SE and NW galaxies.
Based on these features, we further discuss the Dragonfly galaxy as a high-redshift analog to a local late-stage merger, the famous Antennae galaxy (NGC 4038/4039) that has a separation of 6.6 kpc and shows a beads-on-a-string morphology of the molecular gas distribution \citep{1983MNRAS.203...31E,2014ApJ...795..156W}.

In a beads-on-a-string scenario, the filamentary structure forms from the tidal effects ascribed to the recent pericentric passage, as suggested for the Antennae galaxy \citep{2012ApJ...760L..25E}.
Each clump in \#Tidal can be treated as a molecular filament, which is the string, that extends over $\sim400$ pc in the projected plane.
Each filament may embed two or more supergiant molecular clouds (SGMCs) at a 100 pc scale, namely the beads, that host star clusters.
Such beads and strings are reproduced in the high-resolution hydrodynamic simulations of a major merger of disc galaxies with a total mass of $10^8\ \msun$ \citep{2010ApJ...720L.149T}.
They are argued to be the result of the gas turbulent motions and will gain more masses through interactions between the merging pairs.

Shocks originating from galaxy-galaxy collisions/interactions can result in higher turbulence such that the observed line width is determined by the turbulent motions and kinetic temperature ($\tnr{T}{kin}$) of gas.
In this case, we can estimate the Mach number -- \red{a measure of the velocity dispersion of the molecular cloud} -- following \citet{2016ApJ...831...16L}: 
\begin{equation}
    \mathcal{M}=\frac{\sqrt{3}\ \sigma}{0.38\ \kms\ T^{0.5}_{\rm 25K}},
\end{equation}
where $T_{\rm 25K}$ is the kinetic temperature of molecular gas divided by 25 K and $\sigma$ is equal to $\rm FWHM/2.35$.
Assuming a typical $\tnr{T}{kin}=45$ K for the warm molecular gas in high-redshift star-forming galaxies \citep{2021MNRAS.501.3926B}, the Mach number reaches $\sim400$ with $\rm FWHM\sim280\ \kms$ for the line profile extracted from \#Tidal.
This high Mach number is in line with the scenario in which the collisions between cold clouds could be supersonic \citep{1999PhR...321....1S}.
Such a large value is more likely to be a result of the unresolved large-scale structure.
A more reasonable scenario is that, under a `beads-on-a-string' speculation, the observed $\rm FWHM\sim280\ \kms$ can be a composition of several SGMCs with narrower FWHMs and different velocity centers \citep[e.g.,][]{2014ApJ...795..156W}.
Assuming an $\mathrm{FWHM\sim60}\ \kms$ which is a typical value for the SGMCs in the overlap regions of the Antennae galaxy and $\tnr{T}{kin}=45$ K, the resultant Mach number is $\sim80$, in agreement with the values found in molecular clouds at 60 pc scale in mergers and SBGs \citep{2016ApJ...831...16L}.
\red{This scenario requires future observations with similar resolutions and higher sensitivities to confirm.}

\subsection{A radiatively efficient RLAGN}
\label{sec: agn activities}
\subsubsection{A growing SMBH}
\label{sec: eddington}
\red{Apart from starburst activities, the gas inflow accompanied by merger events can also fuel the growth of SMBHs \citep[e.g.,][]{2023ApJ...952..121L}.
Therefore, we wonder whether this RLAGN can be associated with an active BH in a rapid growth phase.}
We, therefore, need to derive the corresponding Eddington ratio.
For the reason discussed in \S\ref{sec:sed fitting}, we use the SED fitting results adopting 44 GHz radio data as a conservative estimate.
The RLAGN host galaxy refers to the NW galaxy throughout the discussions in this section.

AGNs can be classified as radiatively efficient (high accretion rate) and inefficient (often low accretion rate) populations based on whether its Eddington ratio is above $\tnr{\lambda}{Edd}=3\times10^{-2}$ or not, respectively \citep{2008MNRAS.388.1011M}.
The Eddington ratio is defined by $\tnr{\lambda}{Edd}=\tnr{L}{AGN,bol}/\tnr{L}{Edd}$, where $\tnr{L}{Edd}=1.26\times10^{38}\left(\frac{\tnr{M}{BH}}{\msun}\right)$.
Since there exist no obvious observational features of AGN activities in the SE galaxy, such as high-excited molecular gas, molecular outflows, and PSF-dominated morphology in HST images \citep[e.g.,][]{2022ApJ...925..157Z}, the RLAGN that resides in the NW galaxy is treated as the primary contributor to the computed $\tnr{L}{AGN,bol}$.
The Eddington ratios are then $\tnr{\lambda}{Edd}\sim0.07-4$ adopting $M_\text{BH}\sim(0.2-10)\times10^{8}\ \msun$ based on the BH mass estimates presented in \S\ref{sec: bh mass}
These values suggest that the RLAGN is likely to be radiatively efficient and the central SMBH is in a fast growth phase and may even enter the super-Eddington regime.

We also note that this $\tnr{\lambda}{Edd}$ is the most conservative estimate and $\tnr{\lambda}{Edd}$ could be larger by a factor of $\sim3$ if we adopt the fitting including 1.4 GHz data or without radio data.
It should also be noted that the scaling relation is broadly scattered for all populations of AGNs, precipitating inevitable uncertainties in $\tnr{\lambda}{Edd}$.
If our target RLAGN is a radiatively inefficient one ($\tnr{\lambda}{Edd}<0.03$), the SMBH should at least have $M_\text{BH}(=\frac{\tnr{L}{AGN,bol}}{1.26\times(\tnr{\lambda}{Edd}=0.03)\times10^{38}})\geq7\times10^{9}\ \msun$ even adopting $\tnr{L}{AGN,bol}=0.9\times10^{46}\ \ergs$ as the lower limit.
However, \citet{2023A&A...672A.164P} has studied the RLAGN host galaxy properties at $0.3<z<4$ and found statistical consistency in the scaling relations between RLAGNs and other AGNs populations.
This colossal mass will make the NW galaxy an outlier extremely deviating from the scaling relations, and only a few such objects are discovered.
We, therefore, forward our discussions without considering this uttermost possibility and we rely on the radiatively efficient scenario for the RLAGN in the NW galaxy.

\subsubsection{Dragonfly as an Extremely Radio-loud Galaxy}
Defining the ratio of $\tnr{L}{jet}$ to $\tnr{L}{AGN,bol}$ as the intrinsic radio loudness $\tnr{\mathcal{R}}{int}$, we find that $\log\tnr{\mathcal{R}}{int}$ is likely to lie above 0, following $\tnr{L}{jet}$ calculated in \S\ref{sec: jet luminosity} and $\tnr{L}{AGN,bol}\sim0.9\times10^{46}\ergs$ listed in Table.~\ref{tab:fitting result}.
We further calculate the specific black hole accretion rate, which is defined as $\mathrm{sBHAR}=\tnr{L}{AGN,bol}/M_\star\ \mathrm{\ergs\ \msun^{-1}}$ \citep{2021ApJ...921...51I}.
The NW galaxy with $M_\star\sim(3-8)\times10^{10}\ \msun$ corresponds to $\mathrm{sBHAR}\sim(1-3)\times10^{35}\ \mathrm{\ergs\ \msun^{-1}}$.
The obtained values of $\tnr{\mathcal{R}}{int}$ and sBHAR are similar to the average values ($\langle\log\tnr{\mathcal{R}}{int}\rangle=0.0$ and $\langle\log\mathrm{(sBHAR/\ergs\ \msun^{-1}}\rangle=35.3$) of extremely radio-loud galaxies (ERGs) defined by $\log(\tnr{f}{1.4GHz,rest}/\tnr{f}{g\ band,rest})>4$ with $\log\tnr{L}{1.4GHz}>10^{24}$ W Hz$^{-1}$ \citep{2021ApJ...921...51I}.
Therefore, the RLAGN in the NW galaxy is classified as an ERG.

At low redshifts, it has been well established that the powerful radio galaxies with jets are representative of the final evolution stage of AGNs \citep{2008ApJS..175..356H}.
These AGNs host SMBHs surrounded by accretion discs with weak gas inflow, so-called advection-dominated accretion flow (ADAF), which leads to low bolometric luminosity but hard spectrum. 
This accretion state is called as the low/hard state and the accretion disc is in the sub-Eddington phase.
Based on the findings presented here, however, it appears that the Dragonfly galaxy, including ERGs, represents a distinct population, which involves AGNs with rapidly growing SMBHs at high redshifts. 

\subsubsection{Origin of jets from a radiatively efficient SMBH}
\label{sec: jets origin}
The estimates for $\tnr{\lambda}{Edd}$ in \ref{sec: eddington} merely represent a statistical expectation since it is directly related to $\tnr{M}{BH}$.
\red{However, the intrinsic dispersion around the $\tnr{M}{BH}-\tnr{M}{\star}$ scaling relation may still push the SMBH residing in the NW galaxy ($\tnr{\lambda}{Edd}\sim0.07-4$) into the super-Eddington accretion regime ($\tnr{\lambda}{Edd}\gtrsim1$).
In this case, the jets in the Dragonfly galaxy, as well as ERGs, might be launched from a super-Eddington accretion disc \citep[or slim disc;][]{1988ApJ...332..646A} being both optically and geometrically thick.}
The disc thickness prevents the diffusion of magnetic fields, and these jets are powered by large-scale magnetic fields in the innermost region and rotating SMBHs and can have an Eddington-order luminosity.
Simulations of jet-disc connections show that jets can be persistently driven by the Blandford-Znajek mechanism \citep[BZ mechanism;][]{1977MNRAS.179..433B} when the accretion flow is super-Eddington \citep{2014MNRAS.441.3177M,2018ApJ...859...28Q}.

The super-Eddington accretion is short-lived and episodic with a timescale ranging $10^4-10^7$ yr, depending on the replenishment of cold gas from host galaxies \citep{2015ApJ...804..148V}.
This is because the gas accretion timescale is inversely proportional to the mass accretion rate squared \citep{2008bhad.book.....K}.
The accretion rate of a super-Eddington thick disc launching powerful jets will decrease as time evolves and finally enter the standard `thin' disc regime.
Shown by the simulation performed by \citet{2023ApJ...954L..22R}, following this state transition, the jet power shows a rapid decrease by more than three orders of magnitude.
The jets are then incapable of maintaining the steady state of the synchrotron radiation spectrum.
As a result, the jets appear to be switched off and no longer interact with the medium, which is speculated in Paper I \citep{2023MNRAS.522.6123Z}.

From this aspect, RLAGNs showing high accretion rates at high redshifts such as the Dragonfly galaxy may represent an indispensable stage of the galaxy-BH co-evolution, through which BHs rapidly gain their masses.
At the same time, the intense star-forming activities lead to a rapid depletion of the molecular gas in the merger as the merging pairs gradually coalesce.
In the end, the accretion rate drops down and the accretion disc is finally settled in the low/hard state.
Correspondingly, the slowly accreting SMBH then resides in a massive, low-SFR ETG with giant radio lobes, which is the typical picture we observe in the nearby Universe.

Another possibility is established upon RLAGNs being analogs to black hole X-ray binaries (BHBs) and microquasars that launch transient jets.
It is widely argued that jets are likely to be launched, which are called continuous jets when the accretion disc is geometrically thick \citep[e.g.,][]{1998tbha.conf..148N,2001ApJ...548L...9M,2014MNRAS.439..503S,2015MNRAS.453.3213S,2015ASSL..414...45T,2019ARA&A..57..467B}, while geometrically thin discs have low continuous jet-launching probabilities since the magnetic fields will diffuse outward at a rate faster than being dragged inward, incapable of powering the jets \citep{1994MNRAS.267..235L,2012MNRAS.424.2097G}.
However, by modeling jet-disc coupling in BHB systems, \citet{2004MNRAS.355.1105F} found that transient jets are launched as the accretion disc transitions from low/hard to high/soft state.
In the low/hard state, the system launches weak continuous jets.
Then the disc luminosity increases accompanied by the emergence of powerful transient jets, and the disc is in an intermediate state.
When the accretion disc is finally settled in the high/soft state, there will be no jets \citep{2004MNRAS.355.1105F}.
The power of transient jets is argued to correlate with the BH spin, but how come they are correlated remains a mystery.
Observational evidence of such a transitioning phase is limited to BHBs or microquasars for which the state transition timescale is short enough.
State transitions in AGN or quasars, if scaled by the orders of magnitude of central BH masses against BHBs, have a period of $10^4-10^7$ yr \citep{2015ASSL..414...45T}.
As studied in Paper I, the RLAGN may have transient or intermittent activities \citep{2023MNRAS.522.6123Z}.
The synchrotron age, i.e., the lifetime of radio jets, has an order of $\sim10^5$ yr.
Therefore, the galaxy merging results in the fueling of the BH and we are likely to be observing an RLAGN whose accretion has transitioned from a low/hard state to a high/soft one.
\begin{figure*}
\begin{center}
\includegraphics[width=0.8\textwidth]{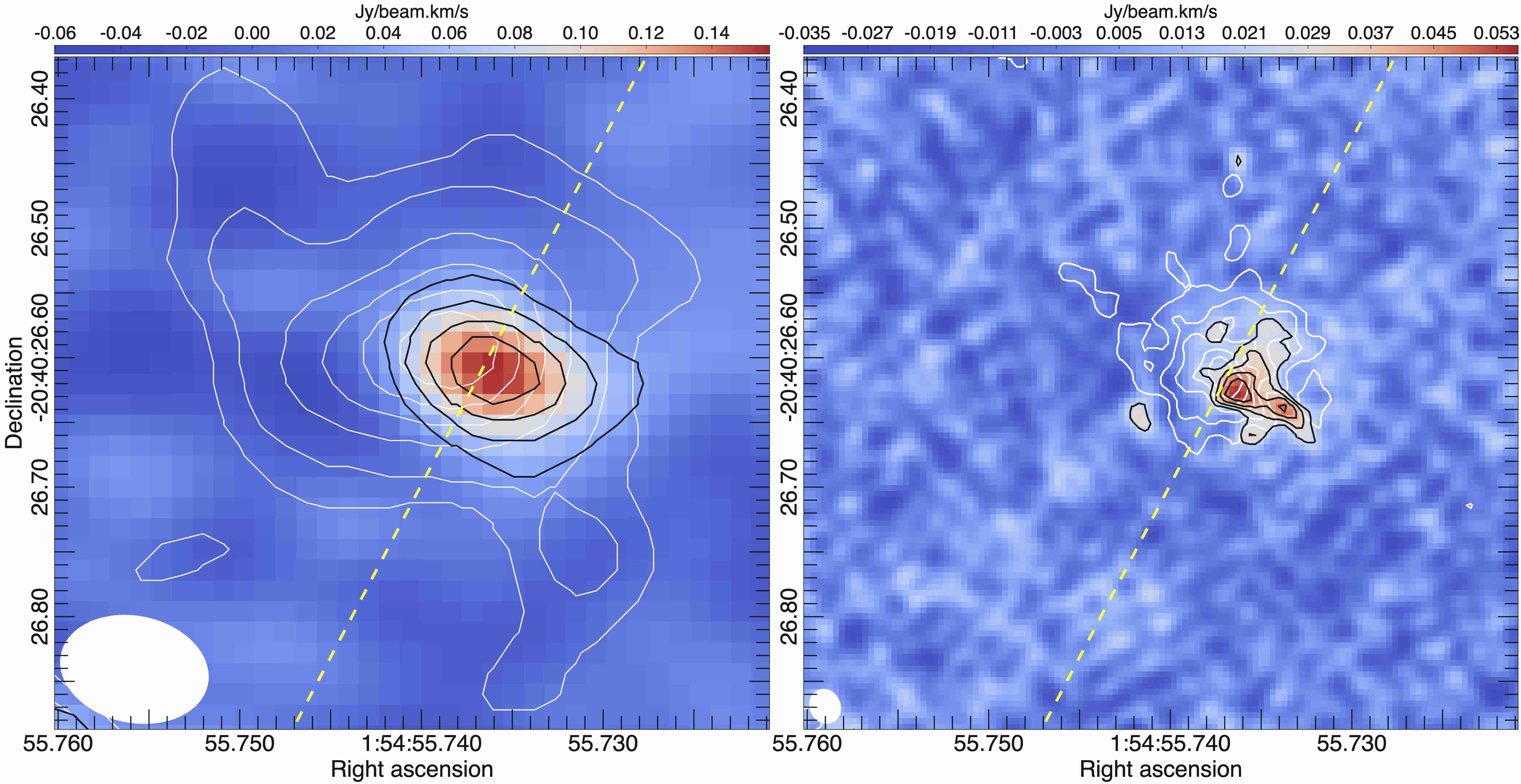}
\caption{
\label{fig:outflow mom0}
The mom0 map integrated from $-500$ to $-280\ \kms$ based on Cycle 4 (left) and 6 (right) observations and indicated by black contours.
Cycle 4 has contour levels of $[3, 5, 7, 9]\times\sigma$ with $\sigma=15\ \mathrm{mJy\ \per{beam}{-1}}$ and Cycle 6 has contour levels of $[3, 4, 5, 6]\times\sigma$ with $\sigma=8\ \mathrm{mJy\ \per{beam}{-1}}$.
\red{The white contours are the mom0 maps integrated from $-280$ to $+250\ \kms$ with contour levels of $[3, 5, 10, 15, 20]\times\sigma$ with $\sigma=18\ \rm m\jbk$ for Cycle 4 and of $[3, 5, 7, 9, 11]\times\sigma$ with $\sigma=11\ \rm m\jbk$ for Cycle 6 observations.}
In these two figures, we manually move the yellow dashed line that represents the jet axis by $0.05\arcsec$ towards the east horizontally given that the astrometry of VLA may lead to a systematic offset of the imaging result.
}
\end{center}
\end{figure*}

\begin{figure*}
\begin{center}
\includegraphics[width=\textwidth]{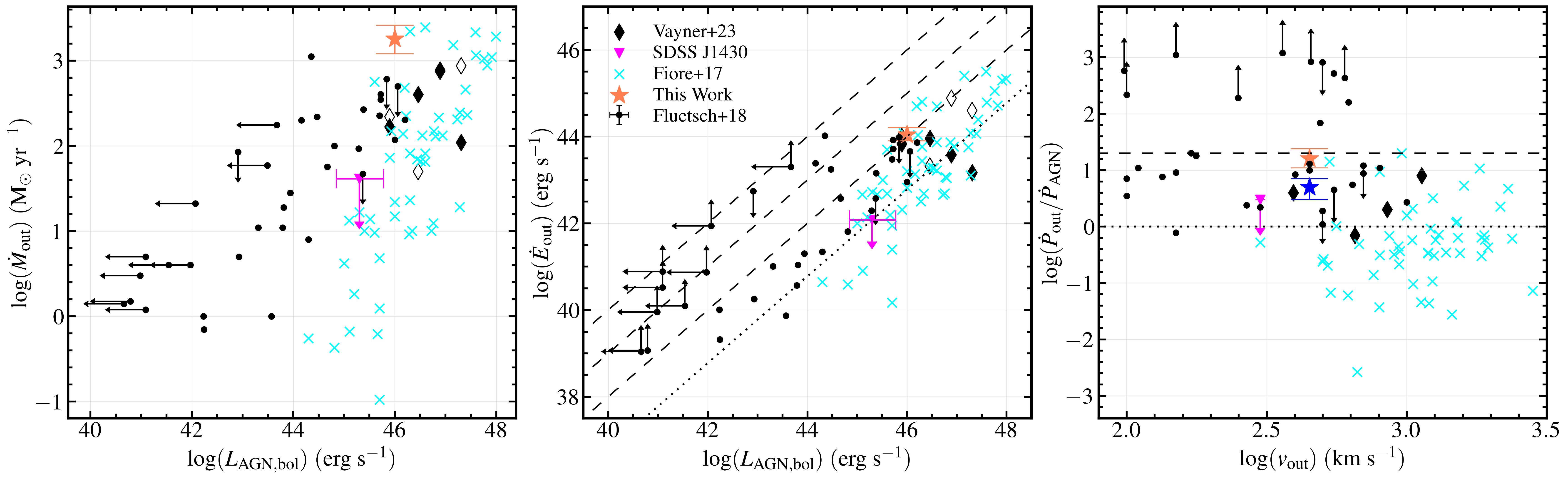}
\caption{
\label{fig:outflow properties}
Molecular outflow properties traced by CO(6-5) of the RLAGN in the NW galaxy are shown by the five-pointed stars. 
\textit{Left}: mass outflow rate as a function of $\tnr{L}{AGN}$.
\textit{Middle}: outflow kinetic power as a function of $\tnr{L}{AGN}$.
\textit{Right}: momentum boost factor as a function of outflow velocity, while the coral stars indicate the calculation adopting $\tnr{L}{AGN,bol}\sim0.9\times10^{46}\ \ergs$, and the blue one adopting $\tnr{L}{AGN,bol}\sim3\times10^{46}\ \ergs$.
Through all panels, dots indicate the molecular outflow of low-redshift AGNs and SFGs from \citet{2019MNRAS.483.4586F}, filled (open) diamonds indicate the molecular (ionized) outflow from radio-loud quasar host galaxies at $z=1.4-2.3$ from \citet{2021ApJ...923...59V}, triangle indicates the upper limit of the molecular outflow in the Type 2 QSO SDSS J1430+1339 at $z=0.085$ \citep{2023A&A...671L..12A}, where $\tnr{L}{AGN,bol}\sim(0.7-6)\times10^{45}\ \ergs$ from \citet{2023A&A...678A.127V} is used, and crosses indicate the ionized outflow from AGNs at $z\sim0.1-3$ \citep{2017A&A...601A.143F} and recomputed based on Eqs. (7-8).
In the middle panel, the dashed lines indicate the thermal-to-kinetic conversion efficiency of 100, 10, and 1\%, while the dotted line indicates 0.06\% for the upper limit of the momentum-driven molecular outflow with $\tnr{\epsilon}{kin}=6\times10^{-4}\left(\frac{\tnr{v}{out}\sim1000\ \kms}{1000\ \kms}\right)$ \citep{2012ApJ...757..136W,2014MNRAS.444.2355C}.
In the right panel, the dashed and dotted lines indicated the expected momentum boost factor for energy- $(\tnr{\dot P}{out}/\tnr{\dot P}{AGN}=20)$ and momentum-driven $(\tnr{\dot P}{out}/\tnr{\dot P}{AGN}=1)$ outflow, respectively.
}
\end{center}
\end{figure*}

\subsection{AGN-driven Outflow}\label{sec: AGN feedback}
\subsubsection{Outflow geometry}
To better investigate the high-velocity component ($\rm NW-1$ in Fig.~\ref{fig:imaging}), we have generated the mom0 map of the NW galaxy integrated over $-500$ to $-280\ \kms$, which is shown in Fig.~\ref{fig:outflow mom0}.
This makes it clear that the high-velocity component is likely to extend along a direction perpendicular to the jet axis.
This component also shows lopsidedness -- only towards the southwest direction.
\red{A similar lopsided outflow cone has been observed in the radio galaxy B2 0258+35 and reproduced in simulations \citep{2022NatAs...6..488M}.
In B2 0258+35, the non-detection of the outflow on the opposite side of the monopolar outflow may be attributed to the obscuration by the CO disc or a much larger clumpiness of the ISM.
This is likely to be the case for the NW galaxy, as evidenced by the diffuse dust continuum -- which could be a result of galaxy interactions (see \S\ref{sec: beads on a string}) -- that extends towards the east direction (panel (c) in Fig.\ref{fig:imaging}).}

What is even more interesting revealed by Cycle 6 observations is that, along the jet axis and towards the northwest, there exists weak but extended CO(6-5) line emission.
\citet{2023ApJ...951...73L} also found a broad, high-velocity component at this position based on combined Cycle 2 and 4 observations tapered to a low resolution of $0.19\arcsec$.
Additionally, at this position, \citet{2023ApJ...951...73L} detected a faint blob in the 237 GHz dust continuum (see their Fig.~8), which appears to be an extended, diffuse emission just above the centroid of the NW galaxy in our Cycle 4 dust continuum image (see panel (a) in Fig.~\ref{fig:imaging}). 
If these observational features in Cycle 4 and 6 observations are real, then they may be indicative of an interaction between the NW jet and CO disc in a few tens pc-scale, which is proposed to explain the high-velocity component coincided with the faint blob by \citet{2023ApJ...951...73L}.
Because of this jet-ISM interaction, the propagation direction of the NW jet is then distorted, resulting in a further misalignment in the inclinations of approaching and receding jets relative to the LOS, as discussed in the Doppler boosting effect in Paper I \citep{2023MNRAS.522.6123Z}.
Numerical simulations show that an inclined jet could strongly couple with the ISM, leading to sub-relativistic and wide-angled outflows \citep{2018MNRAS.479.5544M,2022MNRAS.514.4535T}.
Such a jet-ISM interaction in the innermost region is proposed for the geometry of the narrow-line region of NGC 1068, where splitted CO line emission is observed from the torus to the edge of the circumnuclear disk \citep{2019A&A...632A..61G}.
Due to the low signal-to-noise ratio ($\mathrm{SNR=4}$) of the detection in the current Cycle 6 dataset, this circumnuclear disc-scale jet-ISM interaction is more like a hypothesis and requires future observations with higher sensitivities to confirm.

\subsubsection{Outflow energetics}
We further investigate the outflow properties and energetics by adopting the time-averaged, thin-shell approximation \citep{2005ApJS..160..115R} as a conservative estimate instead of a scenario in which the outflow occupies a spherical volume with a uniform density and has a mass outflow rate larger than the thin-shell scenario by a factor of 3 \citep{2012MNRAS.425L..66M,2017ApJ...836...11G}.
The mass outflow rate is given by 
\begin{equation}
    \tnr{\dot M}{out}=\frac{\tnr{M}{out}\times\tnr{v}{out}}{\tnr{R}{out}},
\end{equation}
where $\tnr{v}{out}\sim450\ \kms$ is the maximum outflow velocity defined as $\Delta\tnr{v}{broad}+\mathrm{FWHM/2}$ and $\Delta\tnr{v}{broad}$ is the line center shift between the narrow and broad components.
We determine the outflow radius by fitting a Gaussian component to the mom0 map (Fig.~\ref{fig:outflow mom0}) integrated from $-500$ to $-280\ \kms$, by which $\tnr{R}{out}\sim0.5$ kpc based on Cycle 6 observations.
The corresponding lower and upper limit, given the molecular gas mass $\tnr{M}{H_2}\sim(1.3-2.8)\times10^9\ \msun$ ($\rm NW-1$ in Table~\ref{tab:lines}), for the mass outflow rate is $1200^{+300}_{-300}\lesssim\tnr{\dot M}{out}\lesssim2600^{+600}_{-600}\ \msun\ \per{yr}{-1}$.
The associated outflow kinetic power is $\tnr{\dot E}{out}=\frac{1}{2}\tnr{\dot M}{out}\times\tnr{v}{out}^2\sim(8-16)\times10^{43}\ \mathrm{erg\ \per{s}{-1}}$ and momentum boost factor 
\begin{equation}
    {\rm momentum\ boost\ factor}=\frac{\tnr{\dot P}{out}}{\tnr{\dot P}{AGN}}=\frac{\tnr{\dot M}{out}\times\tnr{v}{out}}{\tnr{L}{AGN,bol}/c}
\end{equation}
ranges within $11-24$ for $\tnr{L}{AGN,bol}\sim0.9\times10^{46}\ \ergs$ and $3-7$ for $\tnr{L}{AGN,bol}\sim3\times10^{46}\ \ergs$, where $\tnr{\dot P}{AGN}$ is the AGN radiation momentum rate.

We compare these outflow properties with the values in literature in Fig.~\ref{fig:outflow properties}, including a sample of low-redshift AGNs \citep{2019MNRAS.483.4586F}, four radio-loud quasar host galaxies at $z=1.4-2.3$ \citep{2021ApJ...923...59V}, a Type 2 QSO at $z=0.085$ \citep{2023A&A...671L..12A}, and AGNs at $z\sim0.1-3$ \citep{2017A&A...601A.143F}.
The Dragonfly galaxy shows a high molecular outflow rate comparable with the values found in $\tnr{L}{AGN,bol}$-matched AGN host galaxies for both molecular and ionized outflows.
This high rate is a combined result of the compactness ($\sim0.5$ kpc) and massiveness ($\sim2\times10^9\ \msun$).
However, this mass represents merely $\lesssim10$\% of the total molecular gas in the NW galaxy.
Additionally, the starburst activity of the Dragonfly galaxy consumes the molecular gas reservoir at a rate ($\rm SFR\sim2000-3000\ \sfr$) comparable with $\tnr{\dot M}{out}$.
These results suggest that, although the AGN leads to instantaneous negative feedback in the Dragonfly galaxy, it is the long-term and combining effects of AGN and star-forming activities that are responsible for the final quenching, \red{ in line with the ideas proposed in recent simulations \citep[e.g.,][]{2023arXiv231107576B}.}

\subsubsection{Radiative-mode}
\label{sec: radiative-mode}
Based on the estimates of outflow energetics, we first discuss the situation that the outflow stems from the radiative-mode feedback.
\red{In this scenario, the black hole wind drives the expansion of a hot shocked wind bubble that interacts with the host galaxy ISM surrounding the central BH \citep{2003ApJ...596L..27K,2015ARA&A..53..115K}.
If this energy bubble suffers efficient radiative cooling and only transmits the ram pressure to the ISM, the outflow is momentum-driven.
And if the bubble expands adiabatically, the outflow is energy-driven.}

The outflow in the NW galaxy has a kinetic conversion efficiency $\tnr{\epsilon}{kin}=\tnr{\dot E}{out}/\tnr{L}{bol}$ varying within $\sim0.008-0.02$ (see middle panel of Fig.~\ref{fig:outflow properties}).
In a momentum-driven outflow scenario, $\tnr{\epsilon}{kin}$ depends on the outflow velocity by $\tnr{\epsilon}{kin}=6\times10^{-4}\left(\frac{\tnr{v}{out}}{1000\ \kms}\right)$ \citep{2014MNRAS.444.2355C}.
And $\tnr{\epsilon}{kin}$ will be $\sim3\times10^{-4}$ with $\tnr{v}{out}\sim450\ \kms$, which is a factor of 10 smaller than the observed lower limit of $\tnr{\epsilon}{kin}=0.008$.
Therefore, the outflow is unlikely momentum-driven.
The derived $\tnr{\epsilon}{kin}\sim0.008-0.02$ is consistent with $\tnr{\epsilon}{kin}=0.005-0.05$ predicted by simulations for the energy-conserving outflow \citep{2014MNRAS.444.2355C,2015ARA&A..53..115K}.
As shown in the right panel of Fig.~\ref{fig:outflow properties}, the derived $\tnr{\dot P}{out}/\tnr{\dot P}{AGN}\sim11-24$ adopting a lower limit of $\tnr{L}{AGN,bol}\sim0.9\times10^{46}\ \ergs$ supports the energy-driven scenario with the theoretical predictions of $\tnr{\dot P}{out}/\tnr{\dot P}{AGN}\sim20$ \citep{2012ApJ...745L..34Z}.
When $\tnr{L}{AGN,bol}\sim3\times10^{46}\ \ergs$ is considered, we get $\tnr{\dot P}{out}/\tnr{\dot P}{AGN}\sim3-7$, congruent with the prediction of $\tnr{\dot P}{out}/\tnr{\dot P}{AGN}\sim1-10$ if the outflow is directly driven by the radiation pressure with $\tnr{\epsilon}{kin}=0.001-0.01$ \citep{2018MNRAS.476..512I,2018MNRAS.479.2079C}.
This is compatible with our derived $\tnr{\epsilon}{kin}\sim0.008-0.02$.
In conclusion, the radiative-mode AGN feedback is sufficient to drive the massive molecular outflow.

\red{These results yield previous studies of multiphase outflows in $\tnr{L}{AGN,bol}$-matched AGNs \citep[e.g.,][]{2017A&A...601A.143F,2019MNRAS.483.4586F} and QSOs at different redshifts \citep[e.g.,][]{2019A&A...630A..59B,2021MNRAS.504.4445V}, showing that AGN radiative activities are sufficient to explain the outflows observed in systems.
They are also in line with the molecular outflows found in $\tnr{L}{jet}$- and $\sim$$\tnr{L}{AGN,bol}$-matched radio-loud quasars at $z\sim2$ \citep{2021ApJ...923...59V} that reveal energy-conserving outflows.
}

\subsubsection{Jet-mode}
\label{sec: jet-mode}
\red{However, as shown in \S\ref{sec: jet luminosity}, $\tnr{L}{jet}$ is comparable and can even exceed $\tnr{L}{AGN,bol}$, thus we have to consider a second situation that the outflow stems from the jet-mode feedback.}
The radio jets can strongly couple with its host galaxy ISM if the jet has a small angle relative to the disc plane of the host galaxy \citep{2011ApJ...728...29W,2012ApJ...757..136W,2016MNRAS.461..967M}.
The dense molecular clouds can be dispersed away from the jet axis through the ram pressure, forming an expanding cocoon of the molecular gas along the jet path.
The jet flow will channel in various directions because of the porosity of the ISM, resulting in a larger-scale impact on the host galaxy.
However, the outflow component in the NW galaxy extends no more than 0.5 kpc in the projection.
Along with the enhanced velocity dispersion perpendicular to the jet axis, this compactness, as indicated by the simulation performed by \citet{2022MNRAS.516..766M}, may represent the early phase of the jet-driven outflow.

\red{In many previous studies of outflows in nearby jetted systems with radio lobes, the structures of multiphase outflows traced by integrated intensity maps are often found to be elongated along the jet axis \citep{2017A&A...608A..38O,2024MNRAS.527.9322G}.
On the other hand, the [O {\small III}]$\lambda5007$ line velocity widths\footnote{$\rm W70=v_{85\%}-v_{15\%}$ or $\rm W80=v_{90\%}-v_{10\%}$ characterised by the difference between 85th and 15th (or between 90th and 10th) percentile velocities of the line profile}, which is indicative of enhanced velocity dispersion, are found to extend along the direction perpendicular to the jet axis (see \citealt{2021A&A...648A..17V,2022MNRAS.516..766M} and references therein).
In the case of our target, the outflowing molecular gas is perpendicular to the jet axis in both integrated intensity and velocity dispersion maps.}
A similar observational example is the local type 2 QSO SDSS J1430 ($\tnr{L}{AGN,bol}\sim(0.7-6)\times10^{45}$ erg s$^{-1}$; \citealt{2023A&A...678A.127V}) whose radio jets have kinetic power of $\tnr{L}{jet}\sim(1-3)\times10^{43}$ erg s$^{-1}$.
The molecular gas outflow is argued to emanate from the lateral outflow/turbulence constituted of warm dense gas excited by the cocoon of shocked gas \citep{2023A&A...671L..12A}.

The jet-medium coupling efficiency is only $\eta=\tnr{\dot E}{out}/\tnr{L}{jet}\sim0.03$ for the Dragonfly galaxy.
Considering the inevitable large uncertainties in the empirical $\tnr{L}{jet}-\tnr{P}{1.4GHz}$ relation, this value represents an upper limit for $\eta$.
The 1.4 GHz flux density is corrected for the particle acceleration and a direct estimate of the jet power could have an order of magnitude increase $(\gtrsim2\times10^{47}\ergs)$, resulting in an extremely low coupling efficiency of $\eta\sim0.001$.
These results suggest that, if the molecular outflow is the dominant phase, the radio jets are weakly coupled with the ISM.
None the less, adopting $\eta=0.01$ and $\tnr{L}{jet}=2\times10^{46}\ \ergs$, the jets provide an total energy injection of $\sim2\times10^{57}$ erg during its active timescale of $\sim2\times10^5$ yr, comparable with the outflow kinetic energy of $\tnr{E}{out}=\frac{1}{2}\tnr{M}{out}\times v^2_{\rm out}\sim3\times10^{57}$ erg.
Albeit this is no more than an order of magnitude estimate, it further shows that the jets are capable of expelling the cold gas during its short active phase.
Based on the recurrent jet scenario proposed in \citet{2023MNRAS.522.6123Z}, the jets may finally remove all molecular gas from its host galaxy through long-term, episodic bursts.

On the other hand, both the outflow mass and kinetic power could be governed by the ionized rather than the molecular phase, which has been observed in RL quasars \citep{2021ApJ...923...59V}.
\red{For the target with $\log(\tnr{L}{AGN,bol})\sim47\ \ergs$ shown in Fig.~\ref{fig:outflow properties}, the ionized phase has a mass outflow rate larger than that of the molecular phase by almost an order of magnitude and requires more energy to be powered.
Similarly, in HzRGs with both $\tnr{L}{jet}$ and $\tnr{L}{AGN,bol}\gtrsim1\times10^{47}\ \ergs$ at $z\sim2$ \citep{2017A&A...600A.121N}, $\tnr{\dot E}{out}$ of the ionized outflow can be two orders of magnitude larger than those derived for the Dragonfly galaxy.
In addition, as studied by \citet{2023A&A...678A.127V}, although SDSS J1430 is weak in molecular outflow, its ionized outflow may require a hundred per cent jet-ISM coupling efficiency in a jet-mode scenario, while the ionized outflow kinetic power and momentum boost factor agree well with the radiative-mode scenario.}
Actually, the simulation shows that when $\tnr{L}{jet}\sim\tnr{L}{AGN,bol}$, jets can more efficiently shock-ionize gas than AGN radiation even jets are aligned $70^\circ$ relative to the galactic plane \citep{2022MNRAS.511.1622M}.
\red{Therefore, for the Dragonfly where $\tnr{L}{jet}\gtrsim\tnr{L}{AGN,bol}$, observations of ionized gas are then essential to investigate whether and how -- in outflow mass and/or kinetic power -- ionized outflow dominates over the molecular phase.}

All in all, the jet-mode feedback is capable of driving massive molecular outflow within a short timescale without additional energy supply from AGN radiation.
On the other hand, the radiative activity of this RLAGN is adequately responsible for the outflow without the assistance of powerful jets.
\red{This makes HzRGs such as the Dragonfly galaxy a unique AGN population amongst which jet- (characterized by the existence of jets) and radiative-mode (characterized by high $\tnr{L}{AGN,bol}$) feedback should co-exist while their relative importance in shaping host galaxies remain unsettled.
Additionally, unlike the RL quasars studied by \citealt{2021ApJ...923...59V} showing consistent distributions between jet propagation axes and outflows, the fact that the outflow in the NW galaxy is perpendicular to the jet axis differs from the common picture.
Future statistical studies of multiphase outflows in radiatively efficient RLAGNs are required to explore the connections between $\tnr{L}{jet}$, $\tnr{L}{AGN,bol}$, jet ages, and outflow kinematics.}

\section{Conclusion}
In this work, we have studied the molecular gas traced by CO(6-5) line emission in the Dragonfly galaxy, a hyper-luminous infrared, starburst major merger at $z=1.92$ using ALMA and VLA observations.
We have studied the gas kinematics using $^\mathrm{3D}$Barolo and performed SED fitting covering optical-to-radio using CIGALE.
Our major findings are as follows:
\begin{enumerate}[leftmargin=*]

\item The SED fitting using optical-to-44 GHz photometric data gives an $\mathrm{SFR\sim3100\ \sfr}$ averaged over 10 Myr and $\mathrm{SFR\sim600\ \sfr}$ averaged over 100 Myr, implying a recent starburst that contributes to thirty per cent of the total stellar mass.
The stellar mass of the Dragonfly galaxy computed by the new fitting is $\sim9\times10^{10}\ \msun$, which is one-fifth the value computed using a SED template for early-type galaxies in previous studies \citep{2010ApJ...725...36D}.
The fitting including the radio core flux density returns an AGN fraction of 0.1.
If using the flux density at 1.4 GHz originating from radio hotspots and lobes in the SED fitting, the AGN fraction becomes 0.3.
The corresponding AGN bolometric luminosity is $\tnr{L}{AGN,bol}\sim2.9\times10^{46}\ergs$, which is larger than the cases without 1.4 GHz data where $\tnr{L}{AGN,bol}\sim0.9\times10^{46}\ergs$ but consistent with the fitting without radio data.

\item The bulk of the molecular gas in both SE and NW galaxies is concentrated within a radius of $\sim0.8$ kpc.
\red{The NW galaxy shows a simple, circular distribution of the bulk of the molecular gas while some extended, diffuse molecular gas exists due to tidal effects.}
For the SE galaxy, the molecular gas is primarily constituted of two structures, named \#Main and \#Tidal (Fig.~\ref{fig:se_zoomin}). 
One shows a double-peaked line profile characteristic of a rotating structure.
Another one is likely to be associated with tidal effects and has its line center significantly blueshifted by $\sim200\ \kms$
In a `beads-on-a-string' speculation, this structure may embed several supergiant molecular clouds with a Mach number of $\sim80$, in line with the values derived for cold clouds at 60 pc-scale in mergers and starburst galaxies.

\item The gas kinematic modelings of the SE galaxy find a position angle of $130^\circ$, consistent with the orientation of extension of the main structure named \#Main.
The SE galaxy has a rotation velocity to velocity dispersion ratio of at least $\tnr{V}{rot}/\tnr{\sigma}{V}\geq5$, indicative of being rotation-dominated.
The molecular gas in the NW galaxy is highly disturbed because of the AGN activities and possible tidal effects, leading to a loosely constrained inclination angle of $\sim28^\circ$, in line with the value of $\sim24^\circ$ directly derived from the integrated moment map based on the aspect ratio.
The NW galaxy has $\tnr{V}{rot}/\tnr{\sigma}{V}\gtrsim3$ and is therefore likely to be rotating as argued by \citet{2015A&A...584A..99E} and \citet{2023ApJ...951...73L}.

\item  Using the $\tnr{M}{BH}-\tnr{M}{\star}$ scaling relations for X-ray-selected AGNs at high-redshifts, we find $\tnr{M}{BH}\sim(0.2-10)\times10^8 \msun$, corresponding to an Eddington ratio of $\tnr{\lambda}{Edd}\sim0.07-4$.
If the SMBH is accreting at the super-Eddington rate, the powerful jets might be launched from an optically- and geometrically-thick super-Eddington accretion disc.
Given also the young age and possible transient and intermittent activities of this RLAGN, the jets might be transient and launched from an accretion disc in a state transition, either from low/hard to high/soft or from a super-Eddington one to the standard thin disc.

\item The molecular outflow has an outflow rate of $\sim1200^{+300}_{-300}-2600^{+600}_{-600}\ \msun\ \per{yr}{-1}$, comparable with the powerful QSO systems.
If the radiative-mode AGN feedback dominates the outflow, the outflow kinetic power of $\sim(8-16)\times10^{43}\ \mathrm{erg\ \per{s}{-1}}$ and momentum boost factor of $3-7$ adopting $\tnr{L}{AGN,bol}\sim0.9\times10^{46}\ \ergs$ and of $\sim11-24$ adopting $\tnr{L}{AGN,bol}\sim03\times10^{46}\ \ergs$ suggest that the outflow could be energy-conserving or be driven through direct radiation pressure on dust.
If the outflow is a result of the powerful jets, i.e., the jet-mode feedback, the jet-ISM coupling efficiency is $\lesssim0.03$ and could be as low as $\lesssim0.001$.
During its short active timescale, the jets can provide a total energy injection of $\sim2\times10^{57}$ erg with a jet-ISM coupling efficiency of 0.01, which is comparable with the outflow kinetic energy of $\sim3\times10^{57}$ erg.
\red{Based on available observations, which mode dominates the outflow remains mysterious.}

\item  The outflowing molecular gas in the NW galaxy is compact with a radius of $\sim500$ pc in projection and appears to be perpendicular to the jet axis and lopsided towards the blueshifted side of the NW galaxy.
Although it is very massive $(\tnr{M}{H_2}\sim(1.1-2.3)\times10^9\ \msun)$, this mass is only $\sim10$\% of the total cold gas deposited in the NW galaxy, suggesting that the AGN feedback is negative but not responsible for the quenching in the short-term.
However, considering the possible intermittent nature of the powerful jets, the jet-mode feedback may quench the host galaxy through the long-term, episodic bursts of jets that will eventually clear out the cold gas.

\end{enumerate}

\section*{Acknowledgments}
\red{We thank the reviewer for the helpful comments to improve the manuscript.}
We thank the staff in ALMA and NRAO helpdesk for their kind help in data calibration and reduction.
This paper makes use of the following VLA data: VLA/15A-316 and VLA/17B-444.
This paper makes use of the following ALMA data: ADS/JAO.ALMA\#2016.1.01417.S and ADS/JAO.ALMA\#2018.1.00293.S.
Data analysis was carried out on the Multi-wavelength Data Analysis System operated by the Astronomy Data Center (ADC), National Astronomical Observatory of Japan.
\red{This research was supported by a grant from the Hayakawa Satio Fund awarded by the Astronomical Society of Japan.}
AKI, YS, and YF are supported by NAOJ ALMA Scientific Research Grant Numbers 2020-16B.
ALMA is a partnership of ESO (representing its member states), NSF (USA) and NINS (Japan), together with NRC (Canada), MOST and ASIAA (Taiwan), and KASI (Republic of Korea), in cooperation with the Republic of Chile. 
The Joint ALMA Observatory is operated by ESO, AUI/NRAO and NAOJ. 
The National Radio Astronomy Observatory is a facility of the National Science Foundation operated under cooperative agreement by Associated Universities, Inc.

\section*{Data Availability}
The ALMA data used in this work are publicly available at https://almascience.nao.ac.jp/aq/.
The VLA data used in this work are publicly available at https://data.nrao.edu.

\bibliography{dragonfly_co}
\bibliographystyle{mnras}

\appendix
\section{Supplements to SED fitting}
In Table~\ref{tab:photometry}, we list the optical-to-radio photometric data used for the SED fitting in \S\ref{sec:sed fitting} and the references from where these data are collected.
Save for ALMA 1.2 mm and VLA 44 GHz, no spatially resolved photometry is available for the Dragonfly galaxy.

\begin{table}
\caption{Photometric data used for SED fitting}
\label{tab:photometry}
\begin{tabular}{lll}
\hline
Band & Flux denstiy (mJy) & Reference \\
\hline
PAN-STARRS \textit{g}-band & $(4.9\pm0.9)\times10^{-3}$ & 1 \\
PAN-STARRS \textit{r}-band & $(7.1\pm0.1)\times10^{-3}$ & 1 \\
PAN-STARRS \textit{i}-band & $(10\pm1.9)\times10^{-3}$ & 1 \\
DECam $g$-band & $(6.16\pm0.16)\times10^{-3}$ & 2 \\
DECam $r$-band & $(8.50\pm0.21)\times10^{-3}$ & 2 \\
DECam $i$-band & $(10.96\pm0.41)\times10^{-3}$ & 2 \\
DECam $z$-band & $(14.31\pm0.75)\times10^{-3}$ & 2 \\
DECam $Y$-band & $(13.89\pm0.31)\times10^{-3}$ & 2 \\
WISE 3.4 $\micron$ & $(95.1\pm2.4)\times10^{-3}$ & 3 \\
WISE 4.6 $\micron$ & $(147\pm5)\times10^{-3}$ & 3 \\
IRAC 3.6 $\mathrm{\mu m}$ & $(97.4\pm1)\times10^{-3}$ & 4 \\
IRAC 4.5 $\mathrm{\mu m}$ & $(149\pm1)\times10^{-3}$ & 4 \\
IRAC 5.8 $\mathrm{\mu m}$ & $(217\pm4)\times10^{-3}$ & 4 \\
IRAC 8.0 $\mathrm{\mu m}$ & $(403\pm8)\times10^{-3}$ & 4 \\
IRS 16 $\micron$ & $1.58\pm0.1$ & 5 \\
MIPS 24 $\mathrm{\mu m}$ & $3.593\pm0.002$ & 6 \\
PACS 70 $\mathrm{\mu m}$ & $22.6\pm3.5$ & 6 \\
PACS 160 $\mathrm{\mu m}$ & $119\pm9.8$ & 6 \\
SPIRE 250 $\mathrm{\mu m}$ & $105\pm8.6$ & 6 \\
SPIRE 350 $\mathrm{\mu m}$ & $81.3\pm7.3$ & 6 \\
SPIRE 500 $\mathrm{\mu m}$ & $64.4\pm6.8$ & 6 \\
ALMA 1.2 mm & $2.1\pm0.6$ & This work \\
VLA 44 GHz & $0.23\pm0.06$ & 7 \\
VLA 1.4 GHz & $453\pm91$ & 8 \\
\hline
\end{tabular}
\begin{tablenotes}
{\item \raggedright  References. (1) \citet{2016arXiv161205560C} (2) \citet{2018ApJS..239...18A} (3) \citet{2019ApJS..240...30S} (4) \citet{2007ApJS..171..353S} (5) \citet{2010ApJ...725...36D} (6) \citet{2014AA...566A..53D} (7) \citet{2023MNRAS.522.6123Z} (8) \citet{1998AJ....115.1693C}}
\end{tablenotes}
\end{table}

\section{Cycle 6 moments}

\begin{figure*}
\begin{center}
\includegraphics[width=\textwidth]{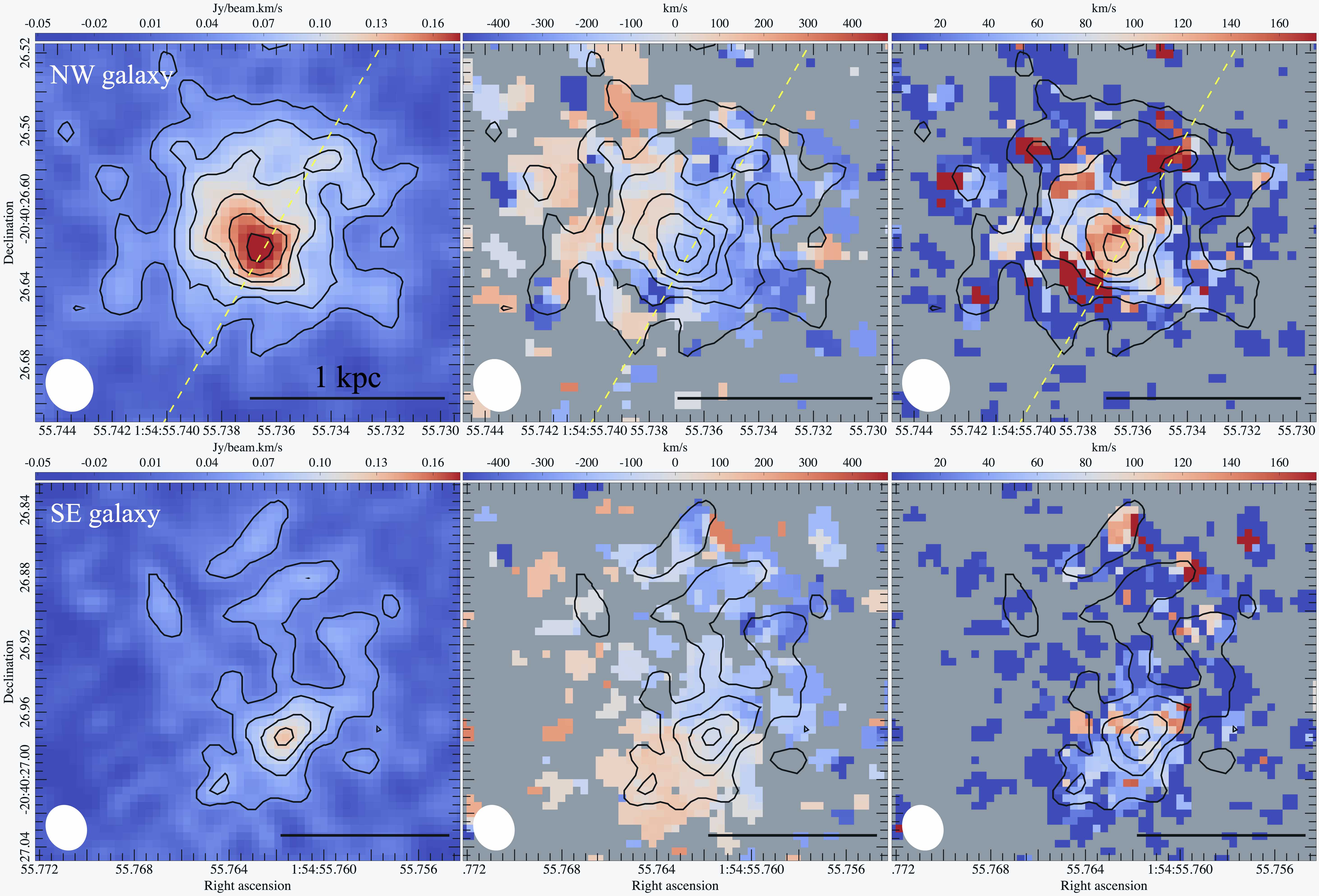}
\caption{
\label{fig:cycle6 moments}
From left to right: the integrated intensity, velocity field, and velocity dispersion maps based on Cycle 6 observations. 
The upper panels show those of the NW galaxy and the lower ones are for the SE galaxy. 
The horizontal black line indicates a 1 kpc physical scale. 
The yellow dashed line indicates the jet propagation direction based
on VLA 44 GHz observation and horizontally translated by $0.025\arcsec$ towards the east.}
\end{center}
\end{figure*}

We show the zoom-in of integrated intensity, velocity field, and velocity dispersion maps of NW and SE galaxies observed in ALMA Cycle 6 in Fig.~\ref{fig:cycle6 moments}.

\section{COMPARISONS WITH HST IMAGING}

We show the HST NICMOS F160W imaging that traces the rest-frame optical continuum of the Dragonfly galaxy, which is overlaid by the ALMA Cycle 4 continuum, Cycle 6 continuum, and C46 CO(6-5) line emission contours.
Due to the poor astrometry of NICMOS imaging, we manually aligned the contours generated from ALMA observations and the HST continuum by matching their centroid.
In panel (c), we label two large-scale structures, the looped tidal arm associated with the NW galaxy and the tidal tail associated with the SE galaxy, which are identified by \citet{2015A&A...584A..99E}.

\begin{figure*}
\begin{center}
\includegraphics[width=\textwidth]{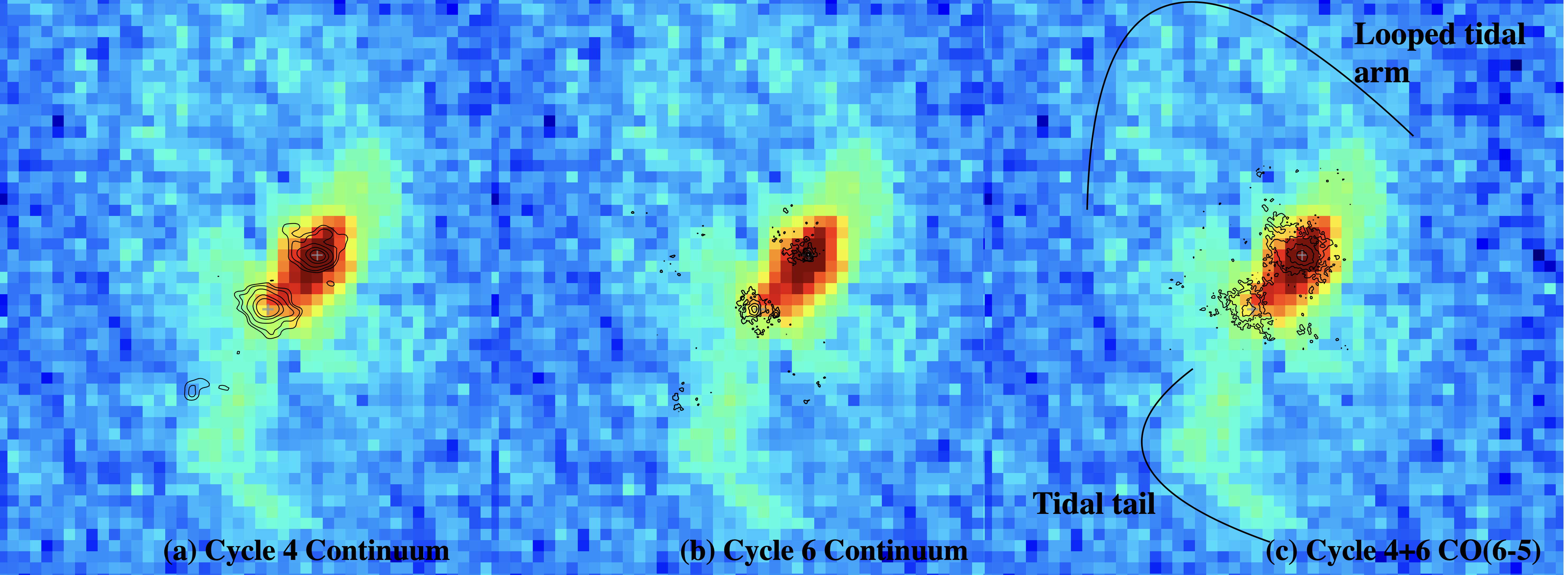}
\caption{
\label{fig:hst imaging}
From left to right, the panels show the HST/NICMOS F160W imaging that traces the rest-frame optical continuum overlaid with ALMA Cycle 4 continuum, Cycle 6 continuum, and Cycle 4+6 CO(6-5) line emission.
The grey crosses indicate the centroid of the NW and SE galaxies.
}
\end{center}
\end{figure*}

\section{Model of ${}^\mathrm{3D}$Barolo}

\begin{figure*}
\begin{center}
\includegraphics[width=0.9\textwidth]{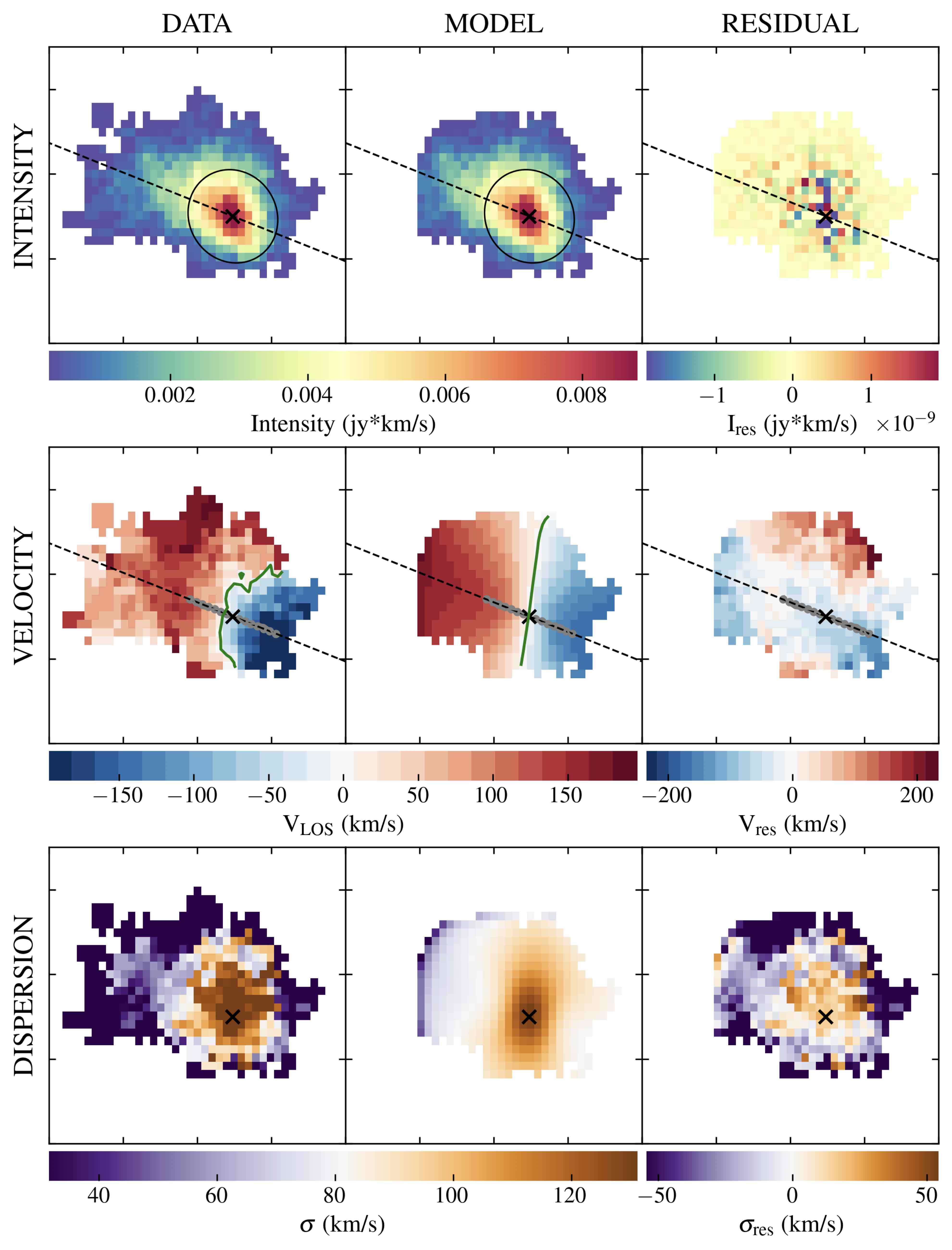}
\caption{
\label{fig:3dbarolo model nw}
The kinematic modeling results using ${}^\mathrm{3D}$Barolo based on C46 observation for the NW galaxy.
From upper to bottom, the panels are integrated intensity, velocity field, and velocity dispersion maps.
From left to right, the panels show the maps directly built from the observational data cubes, the maps built from the modeled data, and the residuals after subtracting the modeled from the observed data.
The dashed line indicates the major axis, the grey points indicate the fitted rings, and the ellipse indicates the outermost ring.
}
\end{center}
\end{figure*}

\begin{figure*}
\begin{center}
\includegraphics[width=0.9\textwidth]{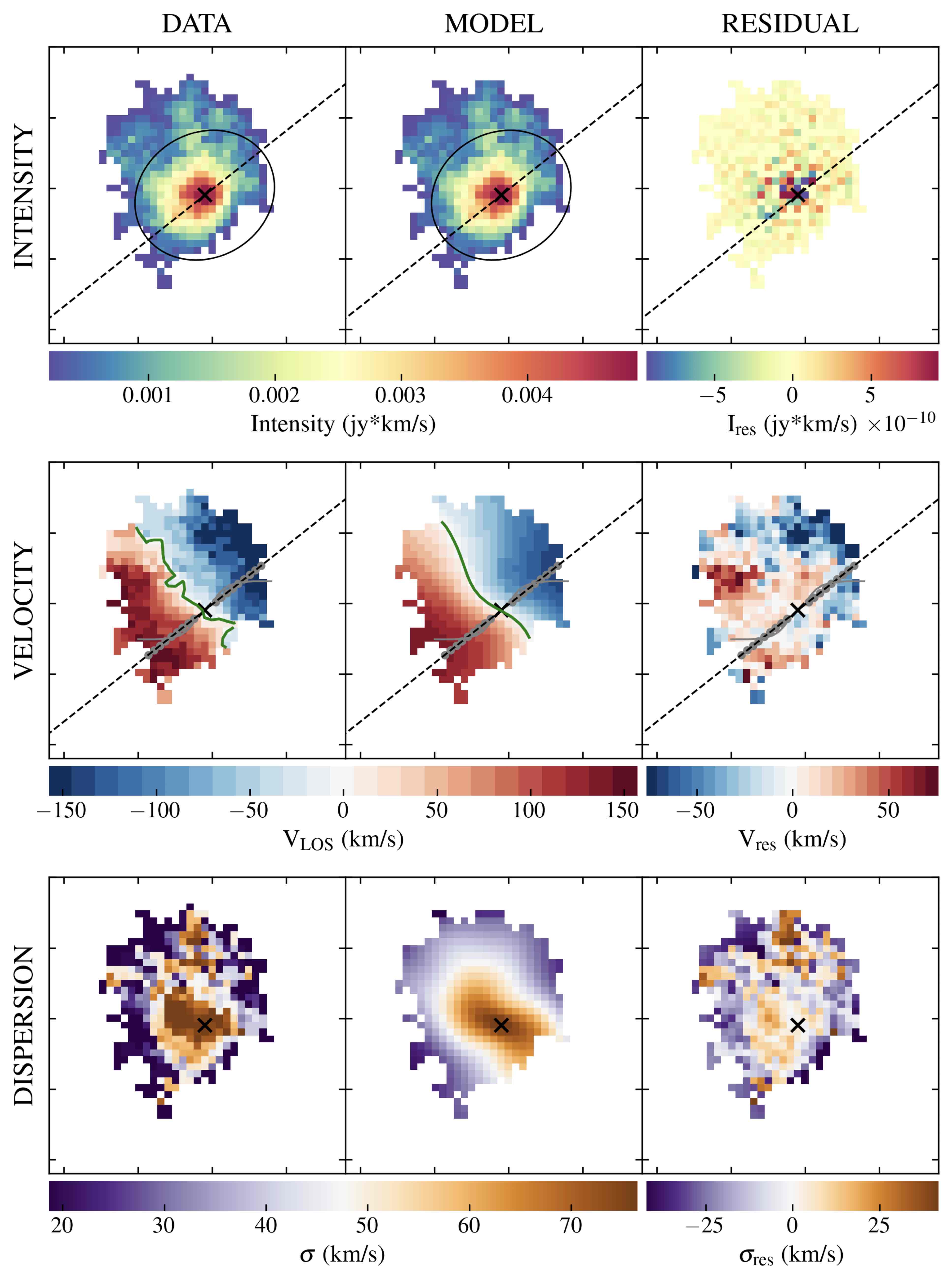}
\caption{
\label{fig:3dbarolo model se}
The kinematic modeling results using ${}^\mathrm{3D}$Barolo based on C46 observation for the SE galaxy.
The panels are the same as above.
}
\end{center}
\end{figure*}

We show the kinematic modelings of NW and SE galaxies using $^{\rm 3D}$Barolo in Fig.~\ref{fig:3dbarolo model nw} and \ref{fig:3dbarolo model se} based on C46 observations.

\bsp	
\label{lastpage}
\end{document}